\documentclass[sigconf, nonacm]{acmart}

\newcommand\vldbdoi{XX.XX/XXX.XX}
\newcommand\vldbpages{XXX-XXX}
\newcommand\vldbvolume{14}
\newcommand\vldbissue{1}
\newcommand\vldbyear{2022}
\newcommand\vldbauthors{\authors}
\newcommand\vldbtitle{\shorttitle} 
\newcommand\vldbavailabilityurl{URL_TO_YOUR_ARTIFACTS}
\newcommand\vldbpagestyle{plain} 

\usepackage{subfigure}

\definecolor{xgreen}{rgb}{0,0.7,0}

\newcommand{\nop}[1]{}
\newtheorem{definition}{Definition}

\newtheorem{example}{Example}

\begin{document}
\title{Learning-Based Data Storage [Vision]\\ (Technical Report)}

\author{Xiang Lian}
\orcid{0000-0001-7681-3807}
\affiliation{%
  \institution{Kent State University}
  \streetaddress{}
  \city{Kent}
  \state{OH 44242, USA}
  \postcode{}
}
\email{xlian@kent.edu}

\author{Xiaofei Zhang}
\affiliation{%
  \institution{The University of Memphis}
  \streetaddress{}
  \city{Memphis}
  \state{TN 38152, USA}
}
\email{xiaofei.zhang@memphis.edu}

\begin{abstract}

\textit{Deep neural network} (DNN) and its variants have been extensively used for a wide spectrum of real-world applications such as image classification, face/speech recognition, fraud detection, and so on. In addition to many important machine learning tasks, as artificial networks emulating the way brain cells function, DNNs also show the capability of storing non-linear relationships between input and output data, which exhibits the potential of storing data via DNNs. We envision a new paradigm of data storage, ``DNN-as-a-Database'', where data are encoded in well-trained machine learning models. Compared with conventional data storage that directly records data in raw formats, learning-based structures (e.g., DNNs) can implicitly encode data pairs of inputs and outputs and compute/materialize actual output data of different resolutions only if input data are provided. This new paradigm can greatly enhance the data security, allow flexible data privacy settings on different levels, achieve low space consumption and fast computation with the acceleration of new hardware (e.g., Diffractive Neural Networks or AI chips), and can be generalized to distributed DNN-based storage/computing.


In this paper, we will propose this novel concept of \textit{learning-based data storage}, which utilizes a learning structure called \textit{learning-based memory unit} (LMU), to store, organize, and retrieve data. As a case study, we use DNNs as the engine in the LMU, and study the data capacity and accuracy of the DNN-based relational tables. Our preliminary experimental results show the feasibility of the learning-based data storage by achieving high (100\%) accuracy of the stored data in the DNN. We also explore and design effective solutions to manage and query the DNN-based relational tables in LMU. Furthermore, we discuss how to generalize our solutions to other data types (e.g., graphs) and environments such as distributed DNN storage/computing. 

\end{abstract}

\maketitle

\pagestyle{\vldbpagestyle}

\begingroup\small\noindent\raggedright\textbf{PVLDB Reference Format:}\\
\vldbauthors. \vldbtitle. arXiv.org, \vldbvolume(\vldbissue): \vldbpages, \vldbyear.\\
\href{https://doi.org/\vldbdoi}{doi:\vldbdoi}
\endgroup
\begingroup
\renewcommand\thefootnote{}\footnote{\noindent
The initial idea of this work on learning-based data storage was born on November 11, 2020. This work is licensed under the Creative Commons BY-NC-ND 4.0 International License. Visit \url{https://creativecommons.org/licenses/by-nc-nd/4.0/} to view a copy of this license. For any use beyond those covered by this license, obtain permission by emailing \href{mailto:info@vldb.org}{info@vldb.org}. Copyright is held by the owner/author(s). Publication rights licensed to the VLDB Endowment. \\
\raggedright Proceedings of the VLDB Endowment, Vol. \vldbvolume, No. \vldbissue\ %
ISSN 2150-8097. \\
\href{https://doi.org/\vldbdoi}{doi:\vldbdoi} \\
}\addtocounter{footnote}{-1}\endgroup


\section{Motivation and Goals}

For the past decades, \textit{deep learning} \cite{SCHMIDHUBER201585,Schmidhuber14} has been widely used for numerous important real applications such as  image classification~\cite{DBLP:conf/nips/KrizhevskySH12}, computer vision~\cite{DBLP:series/acvpr/978-3-319-42998-4}, automatic speech recognition~\cite{8632885}. Examples of many proposed deep learning structures include \textit{deep neural network} (DNN) \cite{Bengio09}, and its variants like \textit{recurrent neural network} (RNN) \cite{Rumelhart86} and \textit{convolutional deep neural network} (CNN) \cite{Goodfellow16}. These learning-based structures (e.g., DNNs) have been shown to produce state-of-the-art results on many machine learning tasks such as classification, regression, and decision making. While DNNs (or other variants of learning-based structures) are trained to essentially capture non-linear relationships between inputs and outputs, the potential of {\bf using DNNs themselves to enable effective and efficient data storage} as well as data management and analysis, similar to the way human brains function, has been underestimated or even ignored in the literature.

In this paper, we {\bf propose a new paradigm of data storage, named ``DNN-as-a-Database''}, which implicitly encodes data pairs of inputs (e.g., resource identifiers) and outputs (e.g., pointers pointing to data resources) via well-trained machine learning models (e.g., DNNs). Compared with conventional data storage that directly records data in raw formats (e.g., relational tables), learning-based structures (e.g., DNNs) do not explicitly release original data. Instead, with the ``DNN-as-a-Database'' paradigm, we use DNN formats, and compute/materialize actual output data (of different granularity) only if input data are provided, which can significantly enhance the data security and allow flexible data privacy settings (e.g., storing attribute value intervals of different resolutions in DNNs). 

Our ``DNN-as-a-Database'' paradigm can be used in scenarios including, but not limited to:
\begin{itemize}

\item {\bf (Distributed Data Storage/Computing)} In the distributed environment, DNNs (rather than original data formats) can be \textit{safely} stored on multiple servers (without disclosing actual data), and only be materialized when input data are provided and computing resources are budgeted. This capability offers the potential to safely store large-scale data on servers and perform secure and parallel data analyses with low communication costs. 

\item {\bf (Secured Data Storage/Computing)} According to different security settings (e.g., whether or not servers and networks are trusted), our DNN-based data storage has the security option of training DNNs over the encrypted inputs/outputs. This option allows data retrieval and query processing without the data decryption and avoids honest-but-curious (HBC) adversaries from breaching the data through networks or servers.

\end{itemize}

Furthermore, with the advent and acceleration of new hardware (e.g., ``programmable'' \textit{Diffractive Neural Network} \cite{Lin18} with miniaturization), DNNs are expected to achieve low space consumption and fast computation (e.g., speed of light), which makes learning-based data storage an attractive option that emulates how human brains memorize data. 

\begin{figure*}[ht]
\centerline{ 
\scalebox{0.62}[0.62]{\includegraphics{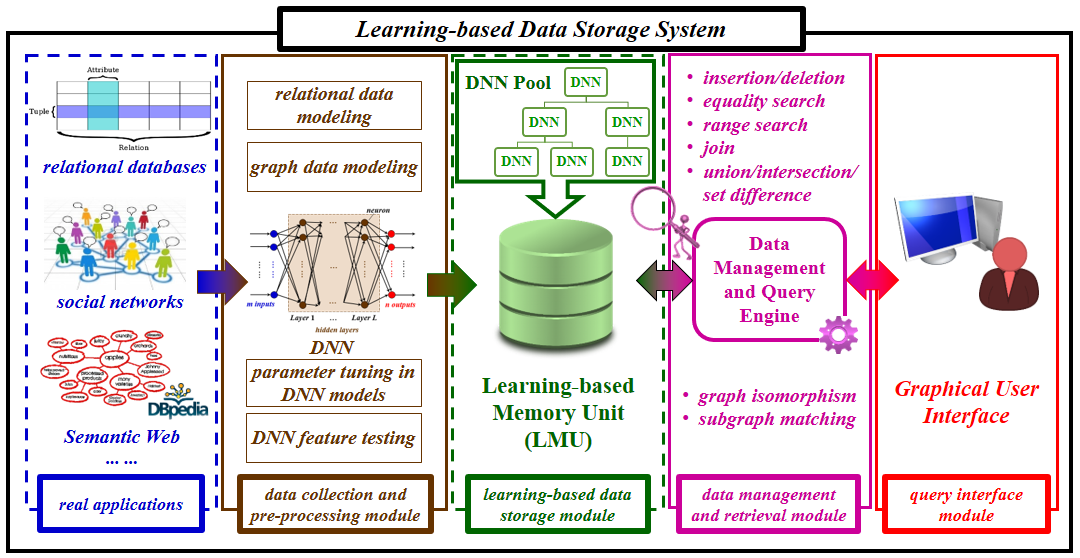}}
}
\caption{\small A Framework for the Learning-Based Data Storage (LDS).}
\label{fig:framework}
\end{figure*}

Therefore, from a novel perspective of memory features of learning-based structures (e.g., DNNs), in this paper, we {\bf envision the potential and feasibility of a novel \textit{learning-based data storage} to utilize learning-based structures such as DNNs to \textit{store}, \textit{manage}, and \textit{analyze} big data}, instead of using DNNs to do classic machine learning tasks such as image classification and object recognition. In order to test features of the learning-based data storage, we will {\bf study and evaluate the accuracy of data that a DNN can encode and store under various parameter settings}. In our preliminary results, we organize a pool of DNNs, as well as auxiliary information (e.g., indexes and/or aggregates), in a so-called \textit{learning-based memory unit} (LMU), which contains a \textit{DNN-tree} structure and can achieve high data capacity and accuracy. 

One major challenge for designing this DNN-based data storage is on how to enable effective maintenance and efficient data analytics over such a DNN-based data storage. As a case study, in this paper, we will {\bf focus on classic structured data, \textit{relational tables}, and transform them to the DNN-based tables in the LMU}. We will {\bf formulate and effectively tackle problems of dynamic maintenance and query processing over DNN-based tables}, including tuple insertions/deletions, attribute updates, projection, selection (e.g., equality/range search), Cartesian product, join, set union, set intersection, and set difference. 


\noindent{\bf A Framework for the Learning-Based Data Storage.} Figure \ref{fig:framework} illustrates a general framework for the \textit{learning-based data storage} (LDS), which consists of four modules: (1) \textit{data collection and pre-processing module}, (2) \textit{learning-based data storage module}, (3) \textit{data management and retrieval module}, and (4) \textit{query interface module}. 

Specifically, the \textit{data collection and pre-processing module} collects real-world application data such as relational tables and tests features of DNN models over these real data (i.e., training data) via parameter tuning in terms of the maximum data capacity and accuracy. The \textit{learning-based data storage module} constructs and organizes a pool of DNNs, as well as their auxiliary information like indexes/aggregates, in the learning-based memory unit (LMU), which stores relational tables and achieves the best data capacity and accuracy. The \textit{data management and retrieval module} provides comprehensive functions/algorithms to maintain and query DNN-based data storage, such as SQL-like queries (e.g., equality/range search, join operator, union/intersection/set difference, etc.) and table updates (e.g., tuple insertions/deletions and attribute updates). Finally, the \textit{query interface module} provides users with a \textit{graphical user interface} (GUI) to manage, query, and visualize tables represented by learning-based structures. These four modules will be seamlessly integrated into a learning-based data storage (LDS) system for users (e.g., data analysts, AI/ML/DM domain experts, or biologists) to analyze the DNN-based data in real applications and/or study the inner mechanisms of learning-based structures for data storage that emulate human memory.

Our proposed DNN-based storage mechanism can be considered as a preliminary work for learning-based data storage. In the future, this work can be generalized to other heterogeneous data (e.g., unstructured data, trees, graphs, etc.) or other computing paradigms (e.g., distributed computing).

The remainder of this paper is organized as follows.  Section \ref{sec:prob_def} formalizes the problem of the DNN-based data storage for relational tables. Section \ref{sec:LMU_design} presents our preliminary results on the LMU design of organizing learning-based structures (i.e., DNNs). Sections \ref{sec:DNN_maintenance} and \ref{sec:DNN_operators} discuss how to maintain and query the DNN-based data storage, respectively. 
Section \ref{sec:discussions} discusses variants of our DNN-based data storage under different security settings. Section \ref{sec:exper} evaluates the performance of our proposed DNN-based data storage through extensive experiments. Section \ref{sec:related_work} reviews previous works on deep neural networks and learning-based data management. Sections \ref{sec:graph_discussions} and \ref{sec:distributed_discussions} discuss how to generalize our learning-based data storage to graph data and distributed environment, respectively. Finally, Section \ref{sec:conclusion} concludes this paper.




\begin{table}[htbp]\small
\caption{Notations and Descriptions}\vspace{-3ex}
\begin{center}
\begin{tabular}{|l||l|}
\hline
\textbf{Notation}&\textbf{Description} \\
\hline\hline
    $T$ (or $R$, $S$) & a relational table containing $N$ tuples\\\hline
    $\mathcal{D}$ (or $\mathcal{D}^R$, $\mathcal{D}^S$) & a DNN-based data storage\\\hline    
    $t_i$ (or $r$, $s$) & a tuple in the relational table\\\hline
    $t_i.id$ & the unique identifier of tuple $t_i$\\\hline    
    $t_i[A_j]$ & the $j$-th attribute of tuple $t_i$ (for $1\leq j\leq d$)\\\hline   
    $m$ & the number of inputs in DNNs\\\hline
    $n$ & the number of outputs in DNNs\\\hline   
    $f$ & the number of neurons for each hidden layer in DNNs\\\hline
    $L$ & the number of hidden layers in DNNs\\\hline
\end{tabular}
\label{tab:notations}
\end{center}
\end{table}

\section{Problem Definition}
\label{sec:prob_def}

Table \ref{tab:notations} depicts the commonly used symbols and their descriptions in this paper.

\subsection{Relational Data Model}

\begin{definition} {\textbf{(Relational Table, $T$)}} A \textit{relational table} $T$ contains a set of $N$ tuples $t_i$, each of which has a unique identifier (e.g., primary key) $t_i.id$ and $d$ attributes $t_i[A_1]$, $t_i[A_2]$, ..., and $t_i[A_d]$, where attributes $t_i[A_j]$ (for $1\leq j\leq d$) have the integer domains. 
\label{def:relational_table}
\end{definition}

In Definition \ref{def:relational_table}, for simplicity, we assume that attributes $t_i[A_j]$ of each tuple $t_i\in T$ are integers, which represent either attribute values or pointers pointing to attribute values of any data types (e.g., texts, numerical values, images, audio, videos, etc.).

\subsection{DNN Model}

\begin{figure}[t!]
\centering 
\scalebox{0.15}[0.15]{\includegraphics{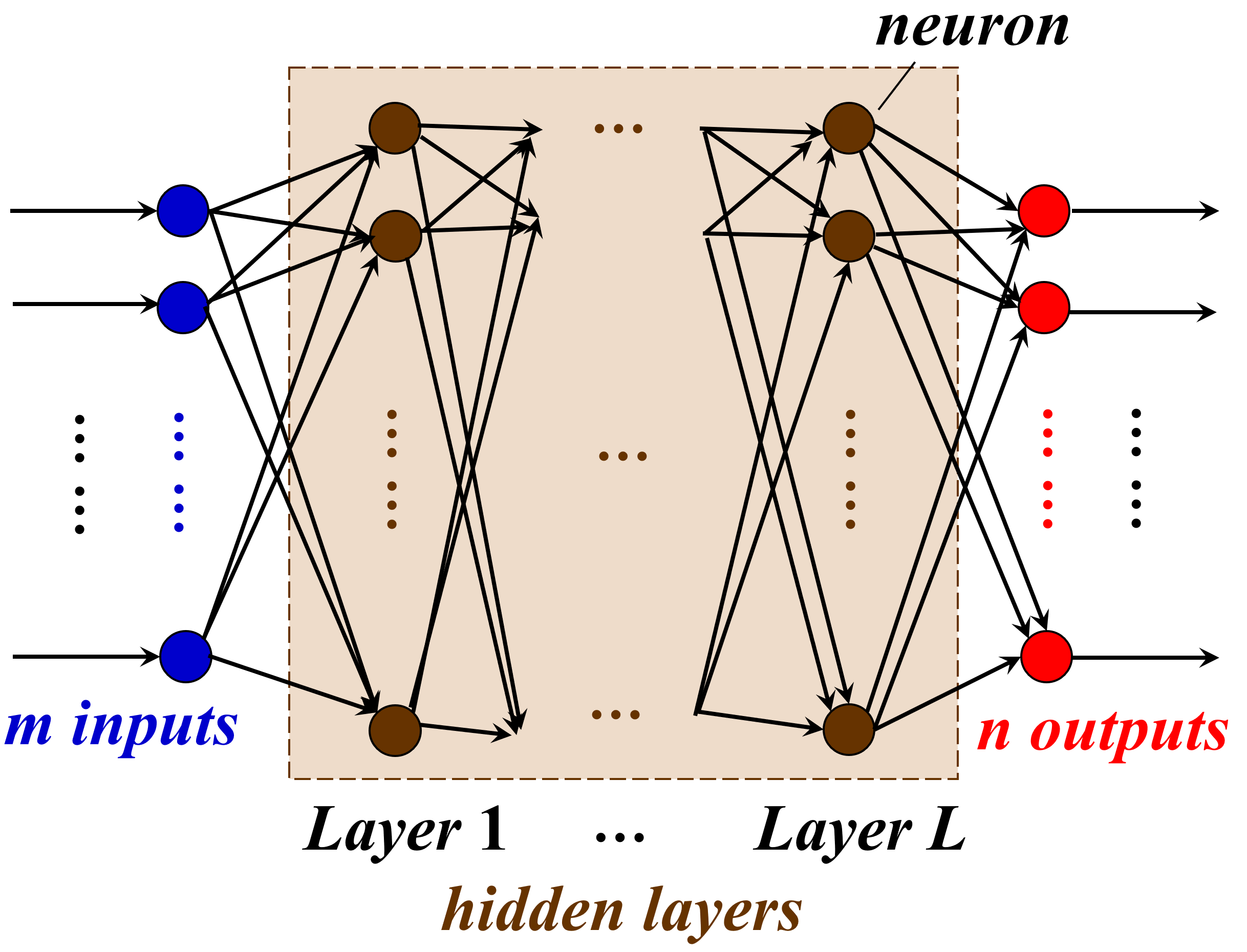}}
\caption{\small Illustration of an $m$-$f[L]$-$n$ Deep Neural Network (DNN).}
\label{fig:DNN}
\end{figure}

As illustrated in Figure \ref{fig:DNN}, a \textit{deep neural network} (DNN) \cite{Bengio09} is an artificial neural network, which contains $m$ inputs, $n$ outputs, and $L$ \textit{hidden layers} ($HL$) (each containing $f$ neurons with \textit{activation functions} such as Sigmoid or ReLU). We denote a DNN as:
$$DNN \sim (in, out),$$ where $in$ and $out$ are vectors of inputs and outputs, respectively. In other words, we have $out = DNN(in)$. 

\noindent {\bf The Space Cost of a DNN.} The space complexity of a DNN is given by $O\big((m+n)\cdot f + f^2\cdot (L-1)\big)$. 

Nevertheless, inspired by recent works on \textit{all-optical deep learning} (e.g., Diffractive Neural Networks \cite{Lin18}), we expect that DNNs may be boosted via 3D-printed hidden layers by consuming small space and achieving the computation with the speed of light (using light signals). We believe that in the near future such new hardware (e.g., ``programmable'' Diffractive Neural Networks with miniaturization, similar to very large-scale integrated circuits, or AI chips) may significantly enhance DNN performance, including both space and computation costs. In particular, the space cost (e.g., for 3D-printed hidden layers) might be reduced to $O(f\cdot L)$.

\begin{figure}[t!]
\centering 
\scalebox{0.28}[0.28]{\includegraphics{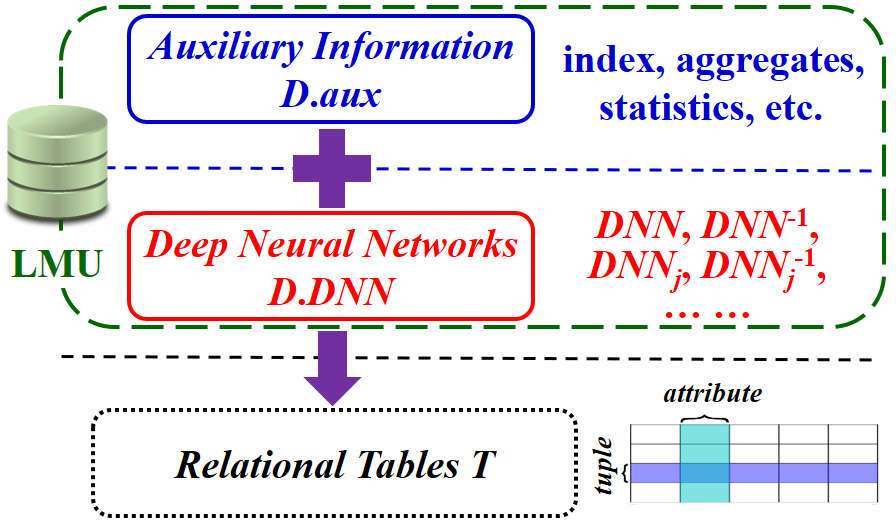}}
\caption{\small Illustration of a Learning-based Memory Unit (LMU), $\mathcal{D}$, Over DNN-based Relational Tables $T$.}
\label{fig:DNN_store}
\end{figure}

In this work, we use the DNN model to train/store multiple input-and-output pairs, where any classic DNN training algorithm can be applied. We would like to consider other structures (e.g., DNN variants) for the learning-based data storage as our future work. 


\subsection{Learning-based Memory Unit for the DNN-based Data Storage}
\label{subsec:DNN_store}

As illustrated in Figure \ref{fig:DNN_store}, we present a \textit{learning-based memory unit} (LMU), denoted as $\mathcal{D}$, to store a relational table $T$, which consists of two components, DNNs $\mathcal{D}.DNN$ and auxiliary information $\mathcal{D}.aux$.

\noindent {\bf Deep Neural Networks, $\mathcal{D}.DNN$.} In $\mathcal{D}.DNN$,  we will train and include the following DNNs to store tuples $t_i$ ($1 \leq i\leq N$) in table $T$: $$DNN \sim ((t_i.id), (t_i[A_1], ..., t_i[A_d])), $$ and $$DNN^{-1} \sim ((t_i[A_1], ..., t_i[A_d]), (t_i.id)),$$ where $DNN^{-1}$ may have multiple outputs (if multiple tuples have the same $d$ attribute values).

For a large-scale relational table, in practice, we can divide tuples into multiple partitions  via vertical/horizontal partitioning, and use multiple pairs of $DNN$ and $DNN^{-1}$ to store individual partitions. Moreover, to implement $DNN$ or $DNN^{-1}$, we will use \textit{DNN-tree} structures as discussed later in Section \ref{sec:LMU_design} to achieve both high data capacity and accuracy. 

\noindent {\bf Auxiliary Information, $\mathcal{D}.aux$.} For auxiliary information $\mathcal{D}.aux$, we will record aggregate data or indexes (e.g., hash files, B$^+$-trees, or multidimensional indexes like grid or R$^*$-trees) over attributes (i.e., inputs of DNNs), for example, the range, $[1, N]$, of tuple identifiers $t_i.id$ for $DNN$.


Note that, to support the maintenance and queries over DNN-based data storage, we will discuss more contents that should be included in $\mathcal{D}.DNN$ and $\mathcal{D}.aux$ later in Section \ref{subsec:selection}.

\underline{\bf \it Sequential Scan.} To sequentially scan the entire table $T$ in the DNN-based data storage $\mathcal{D}$, we first obtain the range of tuple identifiers $t_i.id$ in $T$ from $\mathcal{D}.aux$, for example, $[1, N]$, where $N$ is the number of tuples in $T$. Then, we use the DNN, $DNN$, which takes the input $t_i.id$ from $1$ to $N$ and outputs $N$ $d$-dimensional attribute vectors $(t_i[A_1], ..., t_i[A_d])$, respectively.

\underline{\bf \it Tuple Existence Checking.} Assuming that we have a tuple identifier $t_i.id$, a \textit{tuple existence checking} problem returns true, if this tuple $t_i$ is in $T$; or false, otherwise. Given a tuple identifier $t_i.id$ as the input, we can simply check the existence of $t_i.id$ through $\mathcal{D}.aux$. 

On the other hand, given $d$ attributes $(t_i[A_1], ..., t_i[A_d])$, we can also check whether or not a tuple $t_i$ with these $d$ attribute values exists in $T$ by (1) obtaining a possible tuple identifier $t_i.id = DNN^{-1}(t_i[A_1], ...,$ $t_i[A_d])$, and (2) checking whether or not the possible tuple identifier $t_i.id$ exists in the index of $\mathcal{D}.aux$.




\section{Preliminary Results on the LMU Design}
\label{sec:LMU_design}

In our initial empirical study, we synthetically generate 1,000 distinct tuples, each with 4 random integer attributes (with the domain [1, 1,000]) and a unique tuple identifier (i.e., a key ranging between 1 and 1,000). Below, we outline preliminary results tested on a DNN with 4 hidden layers (128 neurons per layer) using PyTorch~\cite{pytorch}. We report the best accuracy achieved after hyper-parameter tuning. 

\nop{
\begin{itemize}
 \item \noindent {\bf Trial A (Failed):} 4-xx-1 DNN for the regression 

{\bf A.remedy (Failed):} add another level of 1-xx-1 DNN

 \item \noindent {\bf Trial B (Failed):} 4-xx-1 DNN for the classification with 1,000 labels (low accuracy?)

4-xx-1 DNN for classification with 10 labels (intervals) (higher accuracy?)

1-xx-4 DNN $=>$ 1-xx-1 DNN [1 tuple id - xx - 1 attribute]  for classification with 10 labels (intervals) (low accuracy?)

 \item \noindent {\bf Trial C (Successful):}

1-xx-1 DNN $=>$ 4-xx-1 DNN [1 tuple id $=>$ 4 pseudo random numbers (1 tuple id as seed) - xx - 1 attribute] for classification with 100 labels (intervals)

\end{itemize}

}

\begin{figure}[t!]
\centering 
\subfigure[][{\small DNN regression (3.12\% accuracy; Failed)}]{
\scalebox{0.26}[0.26]{\includegraphics{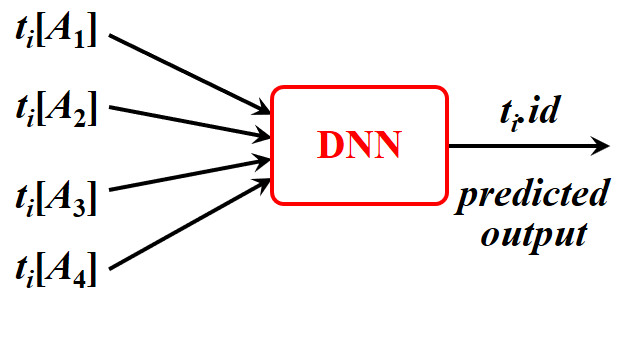}}\label{subfig:trial1}
}%
\subfigure[][{\small classifying tuple ID $t_i.id$  (100\% accuracy; Successful)}]{
\scalebox{0.265}[0.265]{\includegraphics{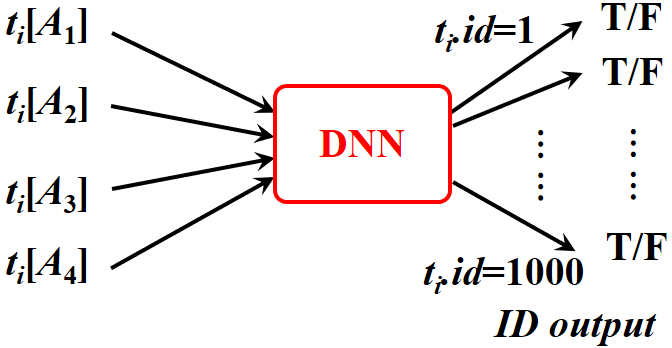}}\label{subfig:trial2}
}\\%
\subfigure[][{\small classifying attribute $t_i[A_j]$  (0.2\% accuracy; Failed)}]{
\scalebox{0.25}[0.25]{\includegraphics{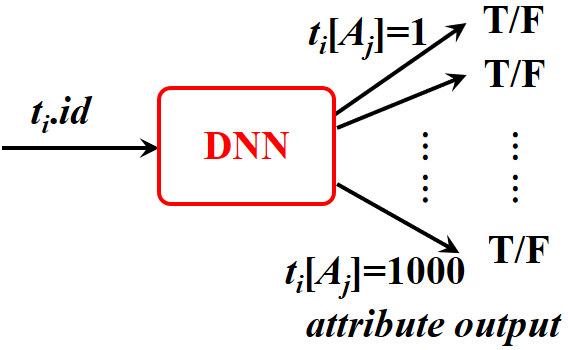}}\label{subfig:trial3}
}%
\subfigure[][{\small classifying attribute $t_i[A_j]$  (100\% accuracy; Successful)}]{
\scalebox{0.25}[0.25]{\includegraphics{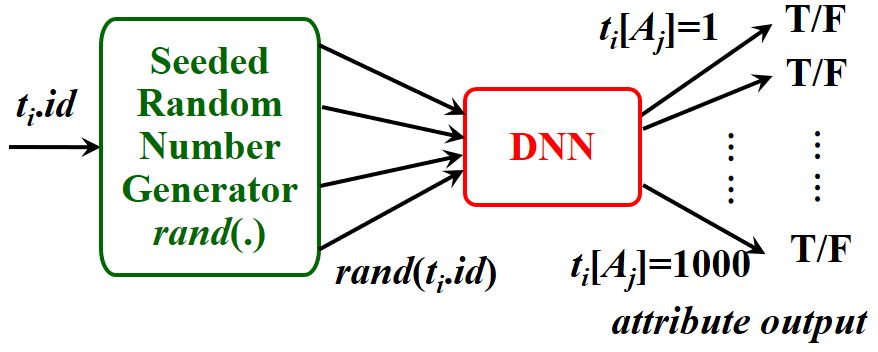}}\label{subfig:trial4}
}
\caption{\small Preliminary Results on the Accuracy of DNN Regression/Classification.}
\label{fig:trials}\vspace{-2ex}
\end{figure}

\begin{itemize}
 \item \noindent {\bf Regression Trial (Failed):} As illustrated in Figure \ref{subfig:trial1}, we tested the DNN regression and used 4 attribute values, $t_i[A_1] \sim t_i[A_4]$, to estimate the tuple identifier $t_i.id$. The predicted tuple identifier usually has a large variance, deviating from the actual one. As a remedy, we tried multi-level structure of DNNs, i.e., using 4 attributes to first predict a value range of the difference between the target ID and current output, and then train another regression model for the records falls in the same variance range. Intuitively,  each range will have fewer records to train/store. However, this trial leads to the best overall accuracy of only 3.12\%, which validates that the regression model via DNNs is not a feasible direction to explore. 




\item \noindent {\bf Classification Trial (Successful):} As illustrated in Figure \ref{subfig:trial2}, by treating each possible tuple identifier as a class, we tested the DNN classification approach and classified 4 attribute values, $t_i[A_1] \sim t_i[A_4]$, to a class (i.e., tuple identifier $t_i.id$). We observed that such a DNN is able to achieve 100\% accuracy. Then, we used a single tuple identifier to predict/classify the attribute values. For simplicity, as illustrated in Figure \ref{subfig:trial3}, we tested a one-on-one mapping from tuple identifier $t_i.id$ to one attribute $t_i[A_j]$. Unfortunately, this model can only achieve an accuracy of 0.2\%, probably due to the input signal being too weak. To fix this issue, we first transformed the tuple identifier $t_i.id$ to a 4-dimensional vector, where vector elements are pseudo random numbers generated by a random number generator, $rand(\cdot)$, with the tuple identifier as the seed. Then, we use this 4D vector as the input of the DNN for classifying/obtaining attribute $t_i[A_j]$. This way, the model can achieve 100\% accuracy in predicting attribute values. By training 4 such DNNs (corresponding to 4 attributes, respectively), we can successfully retrieve the entire record by feeding the tuple identifier $t_i.id$ to the DNNs. 




\end{itemize}

\begin{figure}[t!]
\centerline{
\scalebox{0.28}[0.28]{\includegraphics{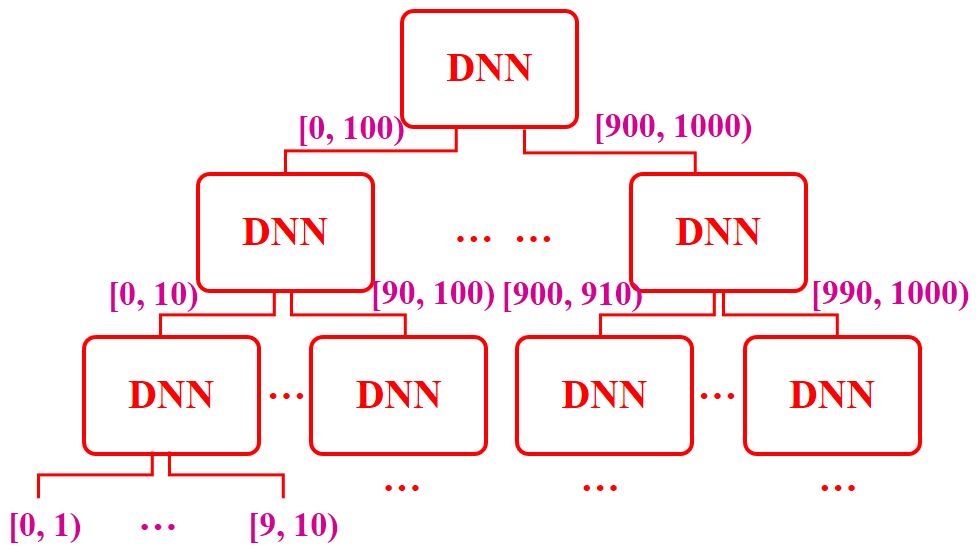}}
}
\caption{\small Illustration of a DNN-Tree.}
\label{fig:DNN-tree}\vspace{-3ex}
\end{figure}

\noindent {\bf Organization of DNNs in the LMU, DNN-Tree.} Inspired by our preliminary results above, we will consider DNNs for classification with labels (classes) corresponding to different resolutions (i.e., attribute value intervals of different lengths). Moreover, since DNNs are not good at predicting too many labels (e.g., 1K or 10K classes) at a time, alternatively, as illustrated in Figure \ref{fig:DNN-tree}, we will organize a pool of DNNs, and build a hierarchical DNN-tree over them, in which each node is a DNN with much fewer labels (e.g., 10 classes), but high (100\%) accuracy. In particular, each child DNN node in the DNN-tree (w.r.t. attribute $t_i[A_j]$) is in the form: $DNN \sim ((t_i.id), (t_i[A_j]))$, which has outputs (i.e., class labels) corresponding to finer value intervals of attribute $A_j$ (e.g., $[90, 100)$), compared with that of its parent DNN node (e.g., $[0, 100)$). This way, at leaf level of the DNN-tree, the precision of the predicted class labels is within $\pm$ 0.5 range, which can accurately predict an integer attribute value (e.g., integer $9$ for interval $[9, 10)$). As a result, we only need to traverse the DNN-tree from the root to a leaf node and incrementally narrow down the range of attribute values from coarse to fine resolutions via DNNs.

Here, as a side product of using the DNN-tree, we can obtain flexible resolutions/intervals of attribute values by controlling the number of levels provided in the DNN-tree, which can be used for \textit{attribute generalization} \cite{Sweene02} in privacy preserving applications.

\begin{figure}[t!]
\centering 
\subfigure[][{\small accuracy vs. table size}]{
\hspace{-3ex}\scalebox{0.2}[0.2]{\includegraphics[angle=270]{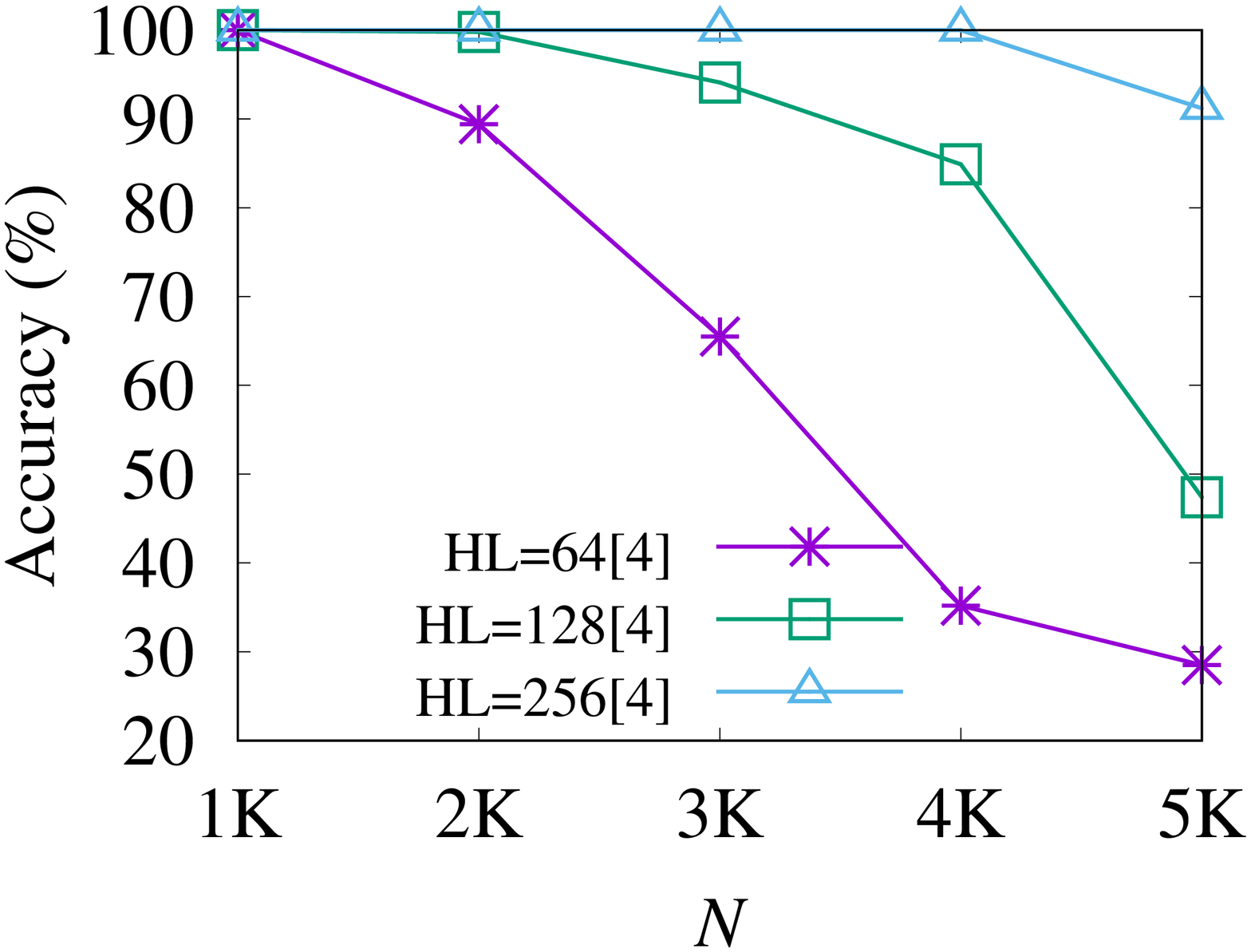}}\label{subfig:cap1}
}%
\subfigure[][{\small max. data capacity vs. \# of classes}]{
\hspace{-4ex}\scalebox{0.2}[0.2]{\includegraphics[angle=270]{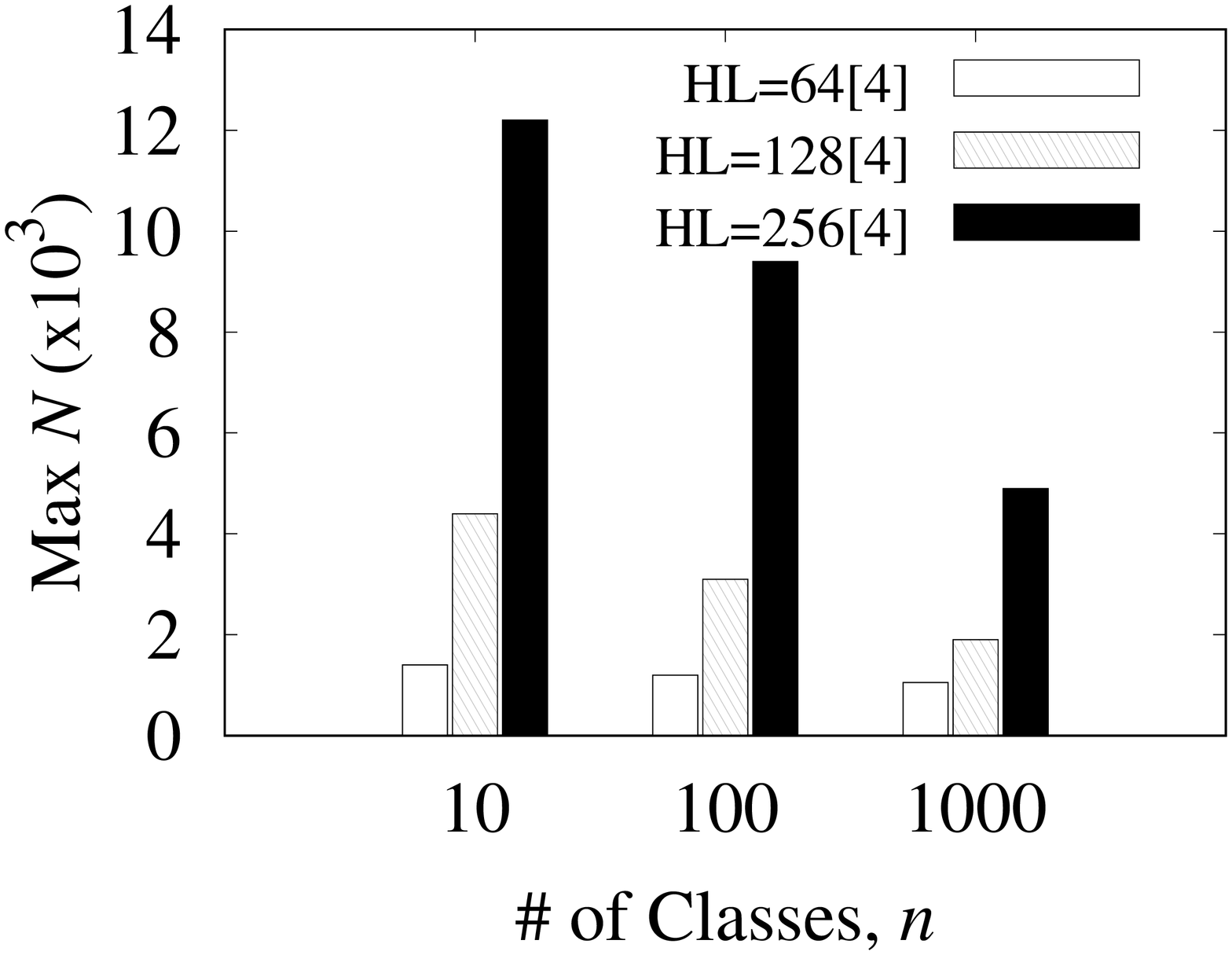}}\label{subfig:cap2}
}%
\caption{\small The DNN Data Accuracy/Capacity Test.}
\label{fig:capacity}
\end{figure}

\noindent {\bf DNN Data Accuracy and Maximum Data Capacity.} 
We also conduct preliminary experiments on a single DNN model ``$DNN \sim (t_i.id,  t_i[A_j])$'' (as given in Figure \ref{subfig:trial4}) and evaluate the accuracy and maximum capacity of data that it can store, by varying the number, $N$, of training/testing tuples or output class number $n$.

As shown in Figure~\ref{fig:capacity}, we use $HL=f[L]$ to denote the configuration of $L$ fully connected hidden layers, where each layer is of size $f$. In our first set of experiments, we empirically validate that larger DNN models can ``memorize'' more tuple data. In Figure~\ref{subfig:cap1}, when $HL=256[4]$, DNN can guarantee 100\% accuracy when the input has 4,000 records, while the other two models (i.e., $64[4]$ and $128[4]$) see significant accuracy drops at this scale of data size. 

In Figure~\ref{subfig:cap2}, we test the maximum number of stored tuples (i.e., max $N$) in the DNN that can deliver 100\% accuracy, for different numbers of output classes $n$. One observation is that when $n$ increases exponentially, $N$ does not drop fast accordingly, especially for DNN of smaller size. For example, for DNN with $HL=128[4]$, the maximum $N$ values with 10, 100, and 1,000 output classes are $4.4K$, $3.1K$, and $1.9K$, respectively.\footnote{In this set of experiments, we evaluate the max $N$ at a step size of 100.} Another intriguing observation is that when we double the layer size $f$ of a DNN, which is equivalent to enlarging the model size by 4 times, the new model's capacity in terms of the max $N$ does not necessarily increase proportionally, especially when the number of classes $n$ is large. It thus raises the optimization opportunity to balance among the DNN model size, accuracy, and capacity given certain (table) data size, which can be used for the parameter tuning of basic DNN units in the DNN-tree.

In subsequent sections, if there is no ambiguity, we will still refer to DNN-trees as DNNs (although there are inner hierarchical structures for DNN-trees) for simplicity.


\section{Maintenance of DNN-based Relational Tables}
\label{sec:DNN_maintenance}

\subsection{Tuple Insertion}

To insert a new tuple $t_{new}$ into the table $T$, we will update the DNN-based data storage $\mathcal{D}$ as follows. 
We will change the range of tuple identifiers in $\mathcal{D}.aux$ from $[1, N]$ to $[1, N+1]$. Moreover, we will re-train two DNNs in $\mathcal{D}.DNN$, that is, $DNN$ and $DNN^{-1}$, over the updated table $T \cup \{t_{new}\}$.

Note that, the tuple insertions can be also accomplished in a batch manner. Either existing DNNs are re-trained by including new tuples, or new DNNs (with tuple identifiers starting from $(N+1)$) are trained over new batch of tuples.

\subsection{Tuple Deletion}

If the table $T$ is append-only, then we do not need to consider the tuple deletion; otherwise, we will design a mechanism to handle the deletion of a tuple $t_{exp}$. Similar to the tuple insertion, we will also update the auxiliary information, such as the index, in $\mathcal{D}.aux$ by removing the tuple identifier $t_{exp}.id$ from $\mathcal{D}.aux$.



\subsection{Attribute Updates}

To update attributes of a tuple $t_i$, we first check whether or not tuple $t_i$ exists in table $T$. If the answer is yes, then we need to train the $DNN$/$DNN^{-1}$ again over the updated table $T \cup \{t_i'\} -\{t_i\}$, where $t_i'$ is the updated tuple of $t_i$.

\vspace{1ex}\noindent {\bf Discussions on Dynamic Maintenance of the DNN-based Data Storage.} From dynamic maintenance of the DNN-based data storage $\mathcal{D}$ above (e.g., tuple insertions and attribute updates), one important issue to improve the performance is to design effective approaches to incrementally and efficiently update DNNs, instead of computing DNNs from scratch. This can be an interesting future research direction. 

\section{Query Operators for DNN-based Relational Tables}
\label{sec:DNN_operators}


\subsection{Projection}

We first consider the projection operator $\Pi_{A_j, ..., A_k}(T)$ as illustrated in Figure \ref{fig:projection}, which projects on attributes $A_j$, ..., and $A_k$ of a relational table $T$. 

In order to enable the projection operator, we will scan table $T$ and return attributes $A_j \sim A_k$ of each tuple $t_i$ in $T$, by using the DNN, $DNN \sim ((t_i.id), (t_i[A_1], ..., t_i[A_d]))$. Specifically, through auxiliary information $\mathcal{D}.aux$, we obtain the range of tuple identifiers $t_i.id$ (e.g., $[1, N]$) as the input of $DNN$, and retrieve the DNN outputs, that is, the projected tuples $(t_i[A_j], ..., t_i[A_k])$.


\begin{figure}[ht]
{\small \sf \begin{tabular}{|l|}\hline 
{\bf select} $t_i[A_j], ..., t_i[A_k]$\\
{\bf from} table $T$\\ \hline
 \end{tabular}
} \caption{\small Projection on Attributes $\Pi_{A_j, ..., A_k}(T)$.}
  \label{fig:projection}
\end{figure}

\subsection{Selection}
\label{subsec:selection}

In this subsection, we consider the selection operator $\sigma_\varphi(T)$ over a relational table $T$, where $\varphi$ is a selection condition. Below, we discuss two basic cases of $\varphi$, that is, $t_i[A_j] = c$ and $c_{min}\leq t_i[A_j] \leq c_{max}$, where $c$, $c_{min}$, and $c_{max}$ are constants.

\noindent{\bf Attribute Equality Search.} We first discuss how to perform the equality search with respect to the $j$-th attribute $A_j$ in table $T$ (represented by the DNN-based data storage $\mathcal{D}$). Figure \ref{fig:equality_search} shows the SQL statement for the equality search, which retrieves those tuples $t_i \in T$ such that the $j$-th attribute $t_i[A_j]$ equals to a constant $c$. 

\begin{figure}[ht]
{\small \sf \begin{tabular}{|l|}\hline 
{\bf select} $t_i.id, t_i[A_1], ..., t_i[A_d]$\\
{\bf from} table $T$\\
{\bf where} $t_i[A_j] = c$\\  \hline
 \end{tabular}
} \caption{\small Selection -- Attribute Equality Search $\sigma_{t_i[A_j] = c}(T)$.}
  \label{fig:equality_search}
\end{figure}

\begin{figure}[t!]
\centering 
\subfigure[][{\small Group By Attribute $A_j$}]{
\scalebox{0.33}[0.33]{\includegraphics{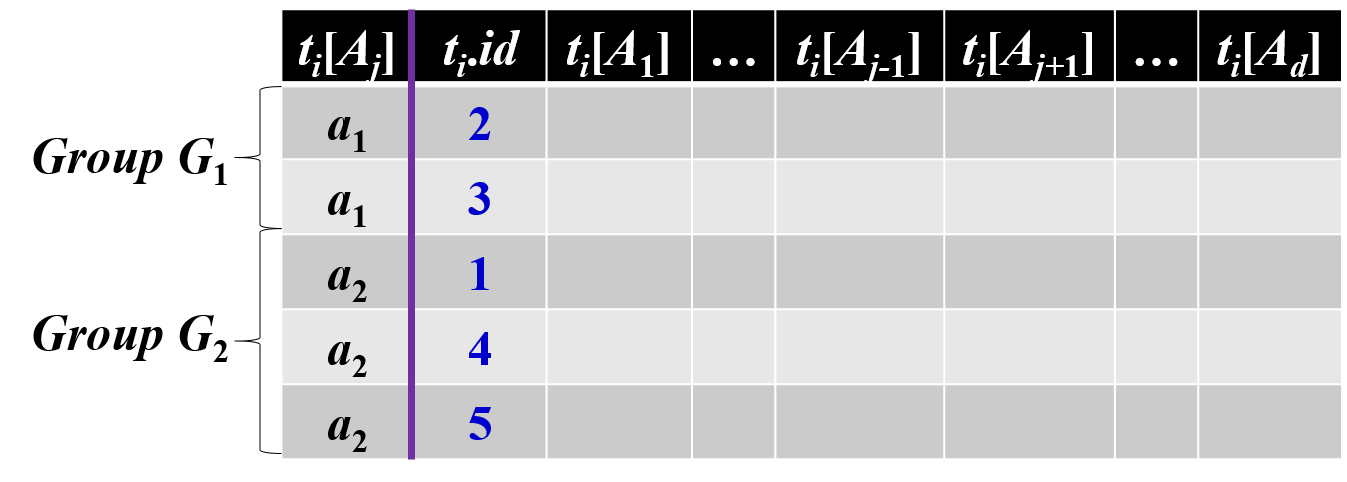}}\label{subfig:equality_DNN1}
}\\%
\subfigure[][{\small Tuple Renaming with New IDs $t_i.id_{new}$}]{
\scalebox{0.33}[0.33]{\includegraphics{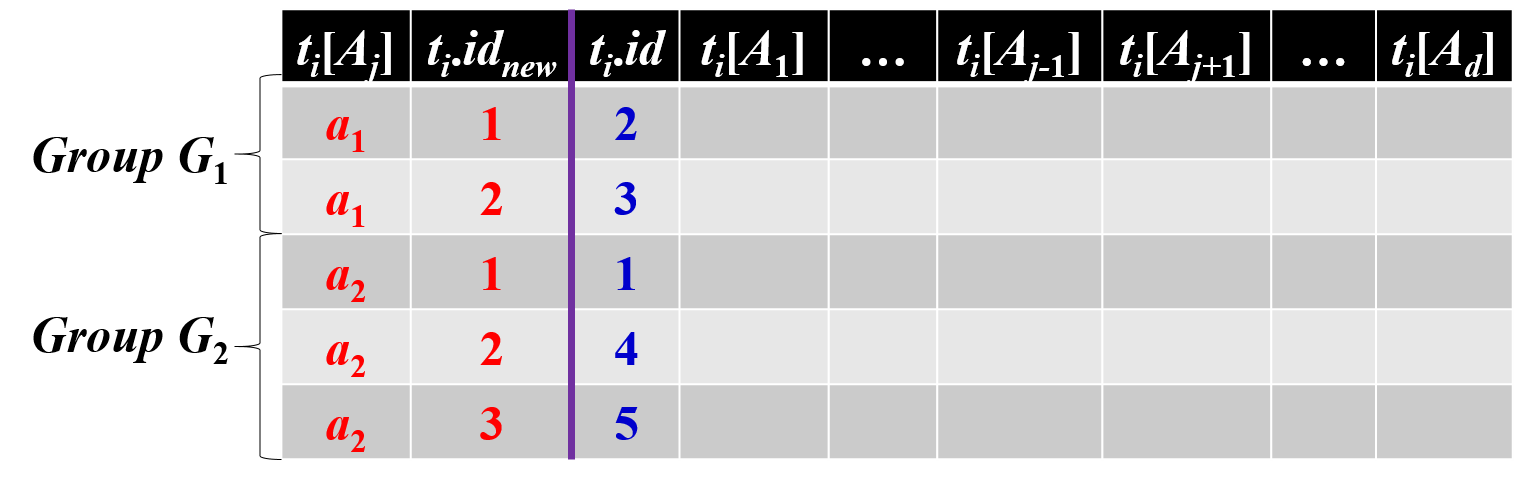}}\label{subfig:equality_DNN2}
}
\caption{\small The Construction of the DNN-based Data Storage for the Equality Search.}
\label{fig:equality_example}
\end{figure}

We will utilize offline pre-computations (e.g., DNN variants) to support online equality search on attribute $A_j$. Specifically, we construct DNN variants in $\mathcal{D}$ as follows. First, we sort all of the tuples $t_i\in T$ based on attribute $A_j$ (in non-descending order), obtaining groups of tuples, $G_1$, $G_2$, ..., and $G_g$ (each group with the same value of attribute $A_j$) as illustrated in Figure \ref{subfig:equality_DNN1}. Each group $G_k$ ($1\leq k\leq g$) is associated with aggregates $(a_k, cnt_k)$, where $a_k$ is the attribute value $A_j$ in group $G_k$ and $cnt_k$ is the size of group $G_k$ (i.e., $|G_k|$). Then, as illustrated in Figure \ref{subfig:equality_DNN2}, each tuple $t_i$ in group $G_k$ is assigned (renamed) with a new unique identifier, $t_i.id_{new}$, within $[1, cnt_k]$.

Next, we build two DNNs, denoted as $DNN_j$ and $DNN_j^{-1}$, to be included in the data storage $\mathcal{D}$ (as illustrated in Figure \ref{fig:DNN_store}):

\noindent $DNN_j \sim ((a_k, t_i.id_{new}), (t_i.id, t_i[A_1], ..., t_i[A_{j-1}], t_i[A_{j+1}], ...,$

\qquad \quad $t_i[A_d])),$ and

\noindent $DNN_j^{-1} \sim ((t_i.id, t_i[A_1], ..., t_i[A_{j-1}], t_i[A_{j+1}], ..., t_i[A_d]),$ 

\qquad \quad $(a_k, t_i.id_{new})).$

Moreover, in auxiliary information $\mathcal{D}.aux$, we will maintain a hash file, $HF_j$, which stores key-value pairs $(a_k, cnt_k)$. 

This way, to perform the equality search in Figure \ref{fig:equality_search}, we first search the hash file $HF_j$ with a search key $c$, and then obtain the matching entry $(c, cnt_k)$. Next, we feed $DNN_j$ with $cnt_k$ input vectors $(c, cnt)$ for $1\leq cnt \leq cnt_k$, and retrieve the corresponding $cnt_k$ tuples (i.e., outputs), respectively, as equality search answers.

\begin{example} Figure \ref{fig:equality_example} illustrates an example of building a DNN-based data storage $\mathcal{D}$ for a table with 5 tuples. We first group all the tuples by attribute $A_j$, and obtain 2 groups $G_1$ and $G_2$ with the same values $a_1$ and $a_2$ of attribute $t_i[A_j]$, respectively (see Figure \ref{subfig:equality_DNN1}). Then, each tuple in the groups is assigned with a new tuple identifier (an integer from 1 to the group size). For example, in Figure \ref{subfig:equality_DNN2}, the first tuple in Group $G_1$ (with old tuple identifier $t_i.id = 2$) is renamed to a new tuple identifier $t_i.id_{new} = 1 
\in [1, 2]$. 

Next, we construct a hash file $HF_j$ in $\mathcal{D}.aux$ over 2 records $(a_1, 2)$ and $(a_2, 3)$ (corresponding to 2 groups $G_1$ and $G_2$, respectively). Following the procedure mentioned above, we also build two DNNs, $DNN_j$ and $DNN_j^{-1}$, in $\mathcal{D}.DNN$ from the table in Figure \ref{subfig:equality_DNN2}. 

For an equality search with query predicate $c=a_1$, we first search $HF_j \in \mathcal{D}.aux$ to find the entry $(a_1, 2)$, and then retrieve the query answers via $DNN_j((a_1, 1))$ and $DNN_j((a_1, 2))$. \qquad $\blacksquare$

\end{example}

\noindent{\bf Attribute Range Search.} Next, we consider the range search on the $j$-th attribute $A_j$ in table $T$ (with the DNN representation). In particular, Figure \ref{fig:range_search} illustrates the SQL query with the range predicate, which obtains tuples $t_i\in T$ with their attributes $t_i[A_j]$ falling into the interval $[c_{min}, c_{max}]$.

\begin{figure}[ht]
{\small \sf \begin{tabular}{|l|}\hline 
{\bf select} $t_i.id, t_i[A_1], ..., t_i[A_d]$\\
{\bf from} table $T$\\
{\bf where} $t_i[A_j] >= c_{min}$ and $t_i[A_j] <= c_{max}$\\  \hline
 \end{tabular}
} \caption{\small Selection -- Attribute Range Search $\sigma_{t_i[A_j] \in [c_{min}, c_{max}]}(T)$.}
  \label{fig:range_search}
\end{figure}

Similar to the equality search, we also need the two DNNs, $DNN_j$ and $DNN_j^{-1}$. To support the range search (rather than the equality search), in $\mathcal{D}.aux$, we will instead maintain a B$^+$-tree, $BT_j$, over attribute $A_j$, with records $(a_k, cnt_k)$ for groups $G_k$.

For any range query with predicate $[c_{min}, c_{max}]$, we first search the B$^+$-tree $BT_j$ for records $(a_k, cnt_k)$ with keys $a_k$ between $c_{min}$ and $c_{max}$. Then, for each retrieved record $(a_k, cnt_k)$, we compute range query answers, $DNN_j(a_k, 1)$, $DNN_j(a_k, 2)$, ..., and $DNN_j(a_k,$ $cnt_k)$.

\underline{\it Discussions on Multidimensional Range Queries.} In Figure \ref{fig:range_search}, the range predicate is on one single attribute $A_j$. In the case that range predicates are specified on multiple attributes (dimensions), one possible way is to estimate the selectivity of the range predicate on each dimension, perform the range search over one B$^+$-tree $BT_j$ with the highest selectivity, and use $DNN_j$ to retrieve candidate tuples and refine them by checking range predicates over other dimensions.

Alternatively, we can also maintain a multidimensional index (e.g., a grid or R$^*$-tree) over all dimensions $A_1 \sim A_d$ in auxiliary information $\mathcal{D}.aux$. Then, cells in the grid or MBRs in leaf nodes of R$^*$-tree can be considered as groups, each of which is associated with a unique group ID $G_k$ (e.g., cell or MBR ID) and the number, $cnt_k$, of tuples in group $G_k$. Each tuple $t_i$ in group $G_k$ can be renamed to $t_i.id_{new}$, where $1\leq t_i.id_{new}\leq cnt_k$. Then, the group-tuple ID pairs $(G_k, t_i.id_{new})$ can be used for constructing the DNNs, $DNN_j$ and $DNN_j^{-1}$, in $\mathcal{D}.DNN$. This way, the resulting DNN-based data storage $\mathcal{D}$ can be used for performing multidimensional range queries (i.e., finding all groups satisfying the range predicates via the multidimensional index, and retrieving candidate answers $DNN_j(G_k, t_i.id_{new})$ for refinement).

\subsection{Cartesian Product}

Figure \ref{fig:cartesian_product} shows the SQL statement of the Cartesian product, $R \times S$, over two tables $R$ and $S$, which returns all possible tuple pairs $(r, s)$ from tables $R$ and $S$, respectively. 

The idea of solving this operator is to sequentially scan both DNN-based tables $R$ and $S$, and enumerate all combinations of tuple pairs. That is, through $\mathcal{D}^R.aux$ and $\mathcal{D}^S.aux$, we obtain ranges, $[1, N_R]$ and $[1, N_S]$, of tuple identifiers, $r.id$ and $s.id$, respectively. Then, for each pair of tuple identifiers $r.id$ and $s.id$ within ranges, we use DNNs, $\mathcal{D}^R.DNN(r.id)$ and $\mathcal{D}^S.DNN(s.id)$, to compute tuples $r$ and $s$ (i.e., outputs of DNNs), respectively, and return $(r, s)$ as the answer.

\begin{figure}[ht]
{\small \sf \begin{tabular}{|l|}\hline 
{\bf select} $r$, $s$\\
{\bf from} tables $R$, $S$\\ \hline
 \end{tabular}
} \caption{\small Cartesian Product $R \times S$.}
  \label{fig:cartesian_product}
\end{figure}

\subsection{Join}
\label{subsec:join}

We next consider the join operator, $R \Join_{r\in R \wedge s\in S \wedge r[A_j] = s[A_l]} S$, over two tables $R$ and $S$, with two DNN-based data storages $\mathcal{D}^R$ and $\mathcal{D}^S$. Figure \ref{fig:join} illustrates the join SQL statement that returns the joining results $(r, s)$ from two tables $R$ and $S$, respectively, where the join condition is given by $r[A_j] = s[A_l]$.

\begin{figure}[ht]
{\small \sf \begin{tabular}{|l|}\hline 
{\bf select} $r$, $s$\\
{\bf from} tables $R$, $S$\\
{\bf where} $r[A_j] = s[A_l]$\\  \hline
 \end{tabular}
} \caption{\small Join $R \Join_{r\in R \wedge s\in S \wedge r[A_j] = s[A_l]} S$.}
  \label{fig:join}
\end{figure}

To enable the join operator, our basic idea is to first perform an \textit{index join or merge} over $\mathcal{D}^R.aux$ and $\mathcal{D}^S.aux$, such as the hash join $HF_j \Join HF_l$, or B$^+$-tree join $BT_j \Join BT_l$, obtain pairs $(r[A_j], cnt_r)$ and $(s[A_l], cnt_s)$, and then retrieve pairwise joining results $(r, s)$ via DNNs (i.e., $\mathcal{D}^R.DNN_j(r[A_j], r.id_{new})$ and $\mathcal{D}^S.DNN_l(s[A_l], s.id_{new})$), where $1\leq r.id_{new} \leq cnt_r$ and $1\leq s.id_{new} \leq cnt_s$.

\subsection{Set Union, Set Intersection, and Set Difference}
\label{subsec:set_operators}

Assuming that tables $R$ and $S$ have the same (compatible) schema, we next discuss the set operators between $R$ and $S$ such as the union, intersection, or set difference, as illustrated in Figure \ref{fig:set_operators}.

For the set operators mentioned above, one fundamental task is to find common tuples between tables $R$ and $S$. In particular, we need to scan one table, say $R$, with a smaller cardinality, by accessing the DNN, $\mathcal{D}^R.DNN(r.id)$, and then check if tuple $\mathcal{D}^R.DNN(r.id)$ appears in the other table $S$ (i.e., whether or not $r.id$ is in $\mathcal{D}^S.aux$). This way, we can obtain common ID pairs $(r.id, s.id)$ from two storages. 

\begin{figure}[ht]
{\small \sf \begin{tabular}{|l|}\hline 
{\bf select} $r$\\
{\bf from} table $R$\\
{\bf UNION} \qquad {\color{xgreen}// {\bf INTERSECT} or {\bf MINUS}}\\
{\bf select} $s$\\
{\bf from} table $S$\\\hline 
 \end{tabular}
} \caption{\small Set Union $R\cup S$, Set Intersection $R\cap S$, and Set Difference $R - S$.}
  \label{fig:set_operators}
\end{figure}


\noindent {\bf Set Union.} For union results $R\cup S$ (with distinct tuples), we will clone two separate DNN storages $\mathcal{D}^R$ and $\mathcal{D}^S$, by excluding common IDs from auxiliary information of one storage (i.e., $\mathcal{D}^R$ or $\mathcal{D}^S$).

\noindent {\bf Set Intersection.} For intersection results $R\cap S$ , we clone one of the two DNN storages, say $\mathcal{D}^R$, and update auxiliary information with common tuple identifiers $r.id$ (i.e., intersecting tuples). 

\noindent {\bf Set Difference.} For the results of the set difference $R-S$, we clone the DNN-based data storage $\mathcal{D}^R$, and update its auxiliary information by excluding those common tuple identifiers $r.id$. 

\nop{
{\color{red}

\subsection{Rename?}

any discussions?

}

}

\nop{

\section{Preliminary Experimental Results}
\label{sec:exper}

data sets: Uniform, Gaussian, Skewed (Zipf), TCP-H?\\

* the number of stored tuples vs. DNN sizes (parameters: dimensionality $d$, DNN parameters such as $L$ and No. of neurons for each hidden layer).\\

* the space cost of the DNN-based data storage vs. the capacity of the DNN-based data storage (plotted chart).\\

* the time cost vs. the data size (sequential scan, tuple existence checking, tuple insertion/deletion/updates, equality/range search, join, union/intersection/set difference).

}

\section{Discussions on Secure and Distributed DNN-based Data Storage}
\label{sec:discussions}

Up to now, we always assume that servers (or networks in the distributed scenario) are all trusted and, thus, input-and-output pairs of DNNs are (integer) plaintexts, for example, for $DNN \sim (in, out)$, $in$ and $out$ are plaintexts without any encryption. In this section, we will discuss how to handle other security settings for our DNN-based data storage in a distributed environment.

Specifically, we will consider 3 cases with different security models (i.e., honest/semi-honest networks or servers):

\begin{itemize}
    \item {\bf Semi-Honest Networks, Semi-Honest Servers (SNS$^2$): } In this SNS$^2$ model, honest-but-curious adversaries may steal data from both networks and servers. To deal with this case, we assume that the data owner (or a trusted third-party) has a (private) symmetric key $K$, and our DNN-based data storage will train DNNs based on the \textit{encrypted} inputs and outputs, that is, $DNN \sim (Encrypt_K(in), Encrypt_K(out))$. 
    
  \quad Therefore, both encrypted inputs and outputs (ciphertexts), $Encrypt_K(in)$ and $Encrypt_K(out)$, will be safely transmitted via semi-honest networks, and semi-honest servers cannot obtain the plaintexts, $in$ and $out$, due to the encryption. After the query issuer obtains the encrypted outputs, $Encrypt_K(out)$, via DNNs from servers, one can compute the decrypted query answer $out = Decrypt_K(Encrypt_K(out))$.

    \item {\bf Semi-Honest Networks, Honest Servers (SNHS): } In this model, networks are not secure, but servers are trusted. Assume that servers have a public key $K_S^+$ and a private key $K_S^-$, and the query issuer has a public key $K_Q^+$ and a private key $K_Q^-$.  Our DNN-based data storage will still train DNNs based on inputs and outputs in plaintext, that is, $DNN \sim (in, out)$. The query issuer will send the encrypted inputs $Encrypt_{K_S^+}(in)$ (ciphertexts with key $K_S^+$) to servers via semi-honest networks, and honest servers will use their private key $K_S^-$ to obtain actual inputs (plaintexts), that is, $in = Encrypt_{K_S^-}(Encrypt_{K_S^+}(in))$. 
    
    \quad After servers compute outputs $out$ via DNNs, they will send $Encrypt_{K_Q^+}(out)$ (with public key $K_Q^+$) back to the query issuer, and the query issuer will use one's private key $K_Q^-$ to obtain the query answer $out$ $=$ $Encrypt_{K_Q^-}$ $(Encrypt_{K_Q^+}$ $(out))$.

    \quad By using asymmetric cryptosystems, we can transmit the encrypted inputs and outputs through semi-honest networks, whereas honest servers perform the DNN computation from inputs to outputs in plaintext. 

    \item {\bf Honest Networks, Semi-Honest Servers (HNS$^2$):} In this model, we will use the same strategy as that in the SNS$^2$ case for the DNN-based data storage/computing. Details are omitted due to similar discussions.

\end{itemize}

\section{Experimental Evaluation for DNN-as-a-Database}
\label{sec:exper}

\subsection{Experimental Setup and Metrics}
To validate our proposed ``DNN-as-a-Database'' paradigm, we conduct preliminary experiments over a prototype solution built with PyTorch 1.13.0 using well-established TPC-H benchmark data sets. All the training and validation are conducted on a server equipped with two CPUs of Intel Xeon Silver 4210, 192 GB Memory, and two GPUs of NVIDIA Quadro RTX 6000, running Ubuntu 20.04, CUDA 11.7. Throughout the experiments, we use GPUs for the model training and CPUs for primitive query operator evaluation. A pooling strategy is adopted for multi-threading, with each thread tying a physical core, 40 threads in total.


\noindent{\bf Data Sets.} We use \textsc{dbgen} of the TPC-H benchmark~\cite{tpch} to generate relational tables, with scale factors of 1 and 10, leading to benchmark tables of total sizes 1 GB and 10 GB, denoted as {\sf TPC-H-F1} and {\sf TPC-H-F10}, respectively.  

\noindent{\bf Default DNN Blocks.} As confirmed by Figure~\ref{fig:capacity}, DNNs can serve as data storage blocks. In our experiments, we adopt a default DNN configuration of 4 hidden layers, 64 neurons per layer, an input vector of size 4, and an output vector of size no larger than 1,000. Our empirical study shows that this setting can achieve a good trade-off among the capacity, evaluation efficiency, and space complexity. More sophisticated hyperparameter tuning for models is beyond the scope of this paper and will be left as our future work. 


\noindent{\bf Metrics.} In subsequent experiments, we aim to address the following questions: 
\begin{itemize}
    \item What are the time/space costs to load relational tables of varied sizes into the DNN-based data storage?
    \item How efficiently can the DNN-based data storage serve primitive relational operators?
\end{itemize}

We report the space cost of our DNN-based data storage, the time cost for offline transforming from relational tables to the DNN-based data storage, and the time cost of SQL query operators (an average over 5 independent runs). 


\subsection{Time/Space Efficiency Evaluation for the DNN-based Data Storage}

\noindent {\bf Space Costs of Relational Tables and DNN-based Data Storage.} Instead of explicitly storing data, DNN-based data storage embeds data into DNN models, i.e., a neural network structure and coefficient weight vectors. It is expected that DNN-based storage would lead to higher space costs, yet our empirical study shows that the space overhead is comparable to raw data size. For two benchmark data sets, {\sf TPC-H-F1} and {\sf TPC-H-F10}, the DNN-based data storage is 2.7x and 2.04x larger than raw relational tables. 

\begin{figure}[t]
\centering 
\subfigure[][{\small {\sf TPC-H-F1}}]{
\hspace{-4ex}\scalebox{0.21}[0.2]{\includegraphics[angle=270]{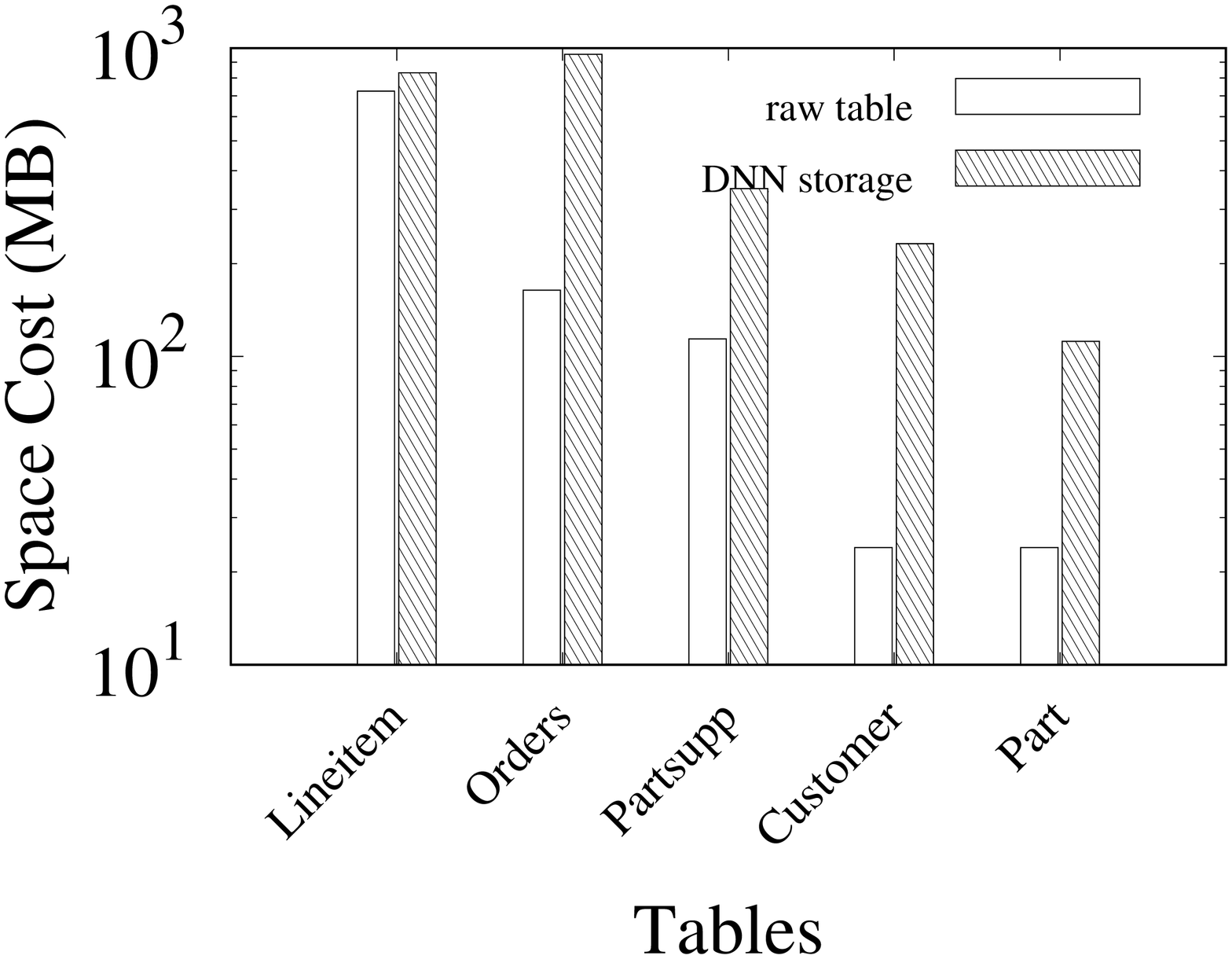}}\label{subfig:table-1}
}%
\subfigure[][{\small {\sf TPC-H-F10}}]{
\hspace{-6ex}\scalebox{0.21}[0.2]{\includegraphics[angle=270]{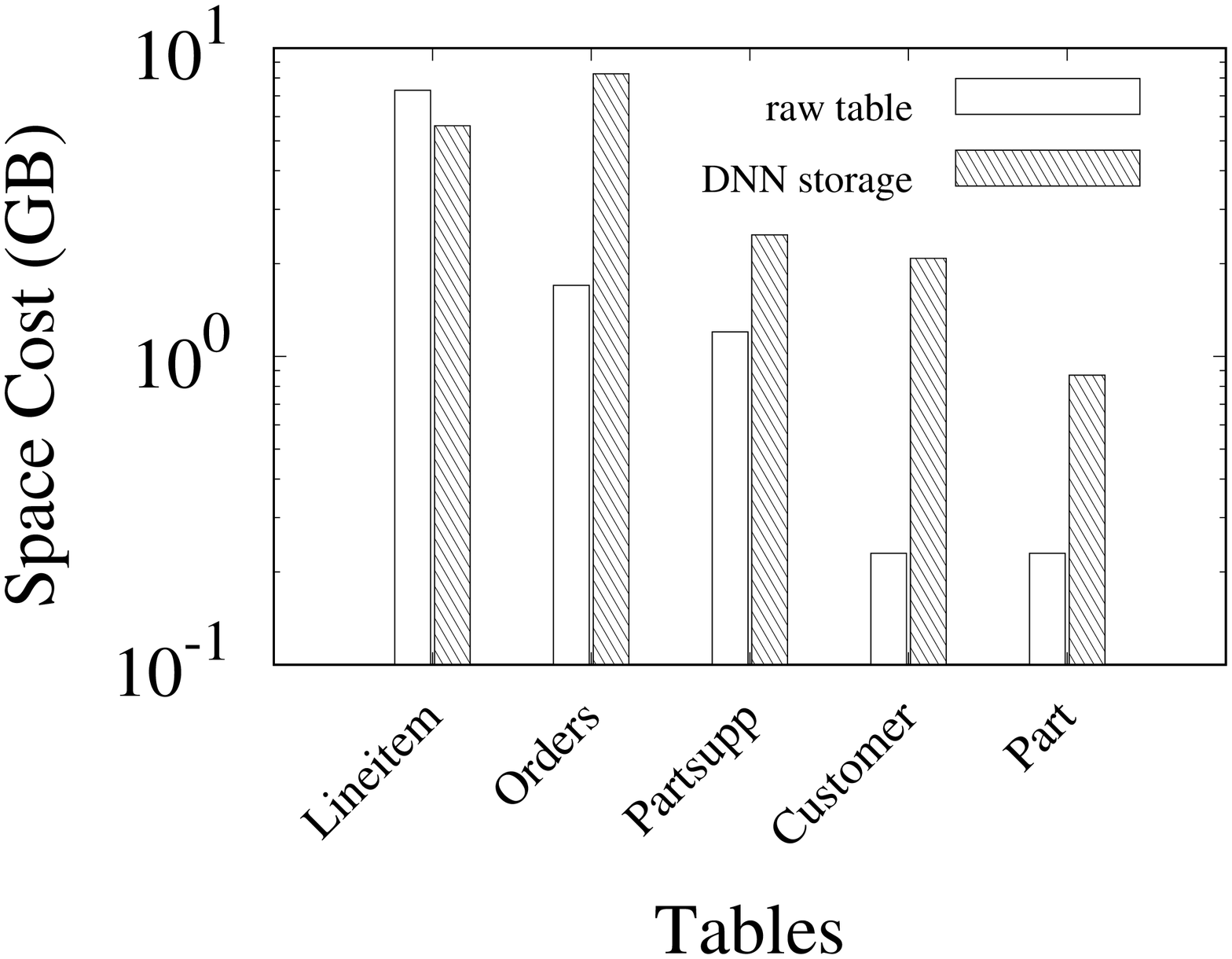}}\label{subfig:table-10}
}%
\caption{\small The Space Costs of Relational Tables vs. DNN-based Data Storage.}
\label{fig:storage}
\end{figure}

Figure~\ref{fig:storage} shows a breakdown of the space cost on different relational tables (note: supplier, nation, and region tables are not reported here, due to small sizes). For the largest table, Lineitem, its DNN-based data storage does not introduce much overhead for {\sf TPC-H-F1} (as shown in Figure \ref{subfig:table-1}); and there is even a size reduction observed for {\sf TPC-H-F10} (as shown in Figure \ref{subfig:table-10}). The reason is that many columns in the ``Lineitem'' table have only a few distinct values (i.e., with many duplicate attribute values). As the DNN-based solution would first group tuples of the same attribute value, it can significantly reduce value duplicates. In contrast, table ``Orders'' has the least number of duplicate attribute values, which leads to as high as 8x space cost for the DNN-based data storage. We observe the similar trend on both benchmarks. Therefore, the space overhead of DNN-based data storage highly depends on the degree of attribute value redundancy in the relational table. 

\begin{figure}[t]
\centering 
\subfigure[][{\small Breakdown of the Table-to-DNN Transformation Time}]{
\hspace{-4ex}\scalebox{0.21}[0.2]{\includegraphics[angle=270]{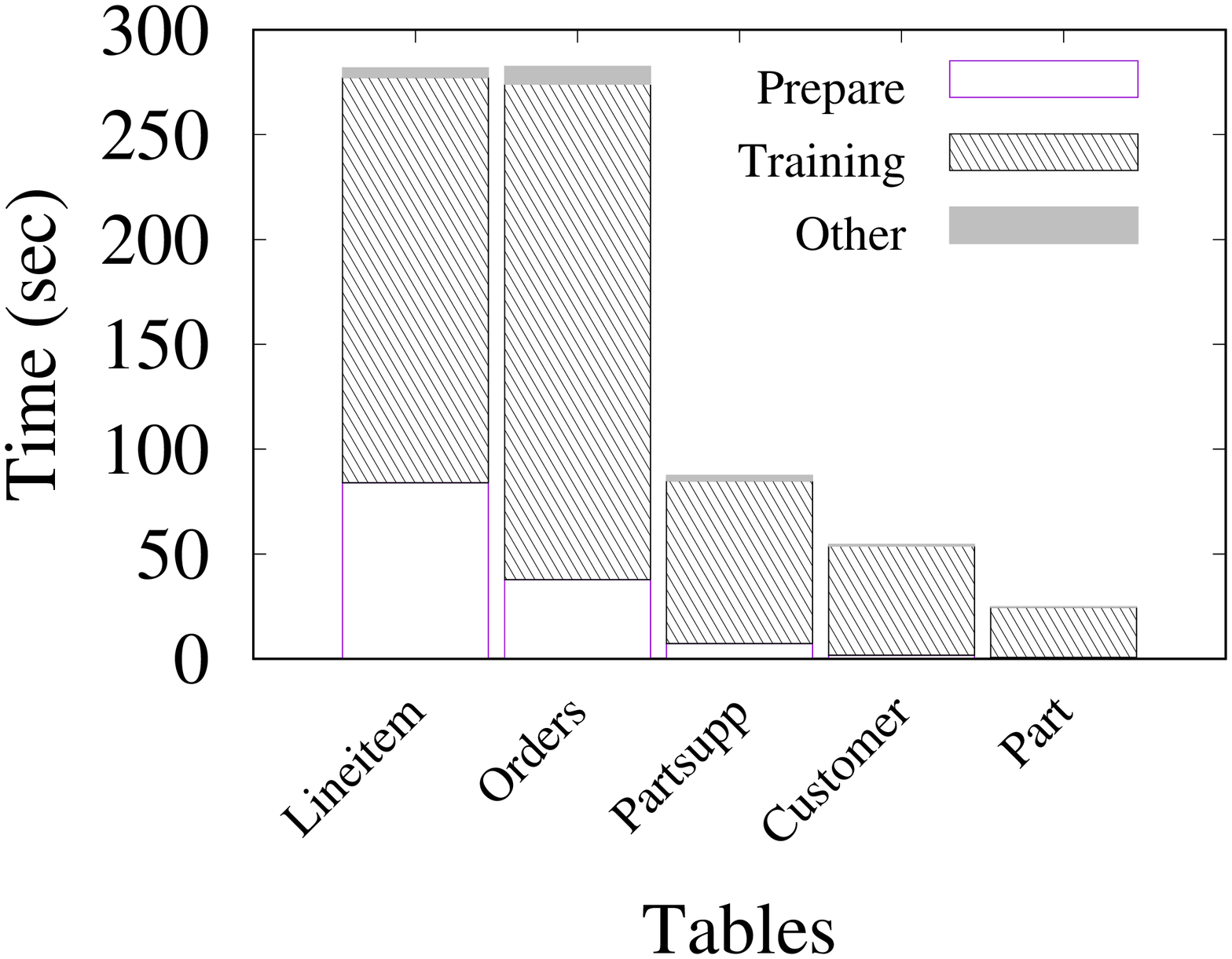}}\label{subfig:table-1-train-time}
}%
\subfigure[][{\small Training Epoch Histogram}]{
\hspace{-6ex}\scalebox{0.21}[0.2]{\includegraphics[angle=270]{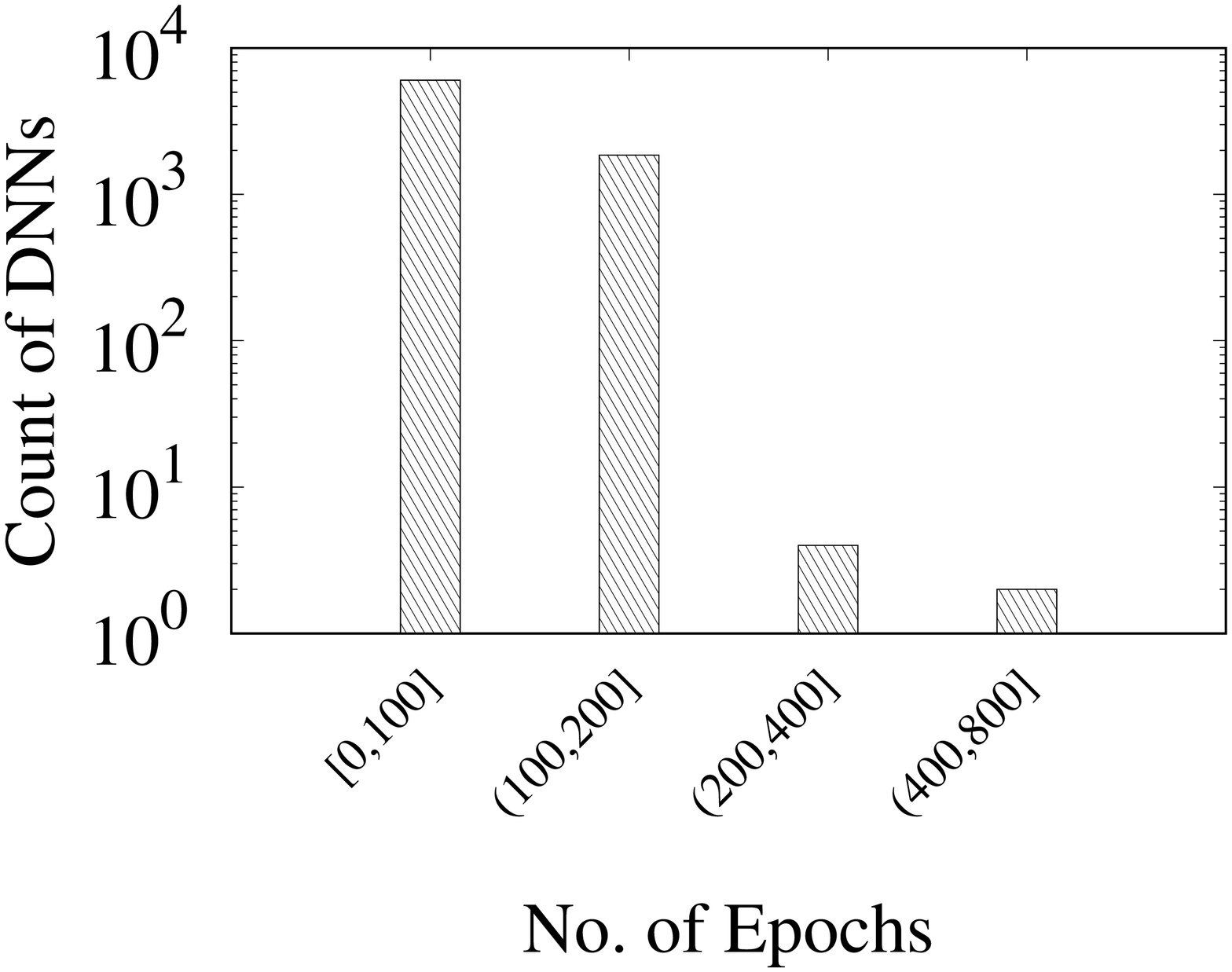}}\label{subfig:epochs}
}%
\caption{\small The Time Cost of Offline Transforming From {\sf TPC-H-F1} to its DNN Format.}
\label{fig:train-time}
\end{figure}

\noindent {\bf The Time Cost of Transforming From Tables to DNN-based Data Storage.} The offline transformation from raw relational tables to their DNN format takes $\sim$775 $secs$ for {\sf TPC-H-F1} and $\sim$5,500 $secs$ for {\sf TPC-H-F10}. In Figure~\ref{subfig:table-1-train-time}, we show a breakdown of time costs to offline transform from raw {\sf TPC-H-F1} tables to their DNN format, including the time costs of the data preparation, training, and others (e.g., I/O, kernel, etc.). The data preparation mainly involves data partitioning (column-wise) and sorting, such that tuples with the same attribute values will be grouped together for supporting effective range search. We further examine the time cost of training default DNN blocks. The training time is a dominating factor in the total transformation time for most tables, which suggests the speedup opportunity given that more GPU power is available. As shown in Figure~\ref{subfig:epochs}, most DNNs (i.e., $\geq$ 99.92\% DNNs) can converge within 200 epochs (with 100\% data accuracy). A similar trend is observed for transforming {\sf TPC-H-F10}. 


\subsection{Primitive SQL Operators Over DNN-based Data Storage}

Next, we examine the efficiency of primitive SQL operators over TPC-H-F1 benchmark tables, including \textsc{selection}, \textsc{projection}, \textsc{range query}, \textsc{Cartesian product}, and \textsc{join}. We compare the query performance of our DNN-based data storage with that of a mature relational database product, PostgreSQL. Note that, we enable parallel query evaluation of PostgreSQL by setting $max\_parallel$ $\_workers$ to 32, $effective\_cache\_size$ to 32GB, $shared\_buffers$ to 4GB, and $work\_mem$ to 256MB.

\begin{figure*}[t!]
\centering 
\subfigure[{\small \textsc{Selection}}]{
\hspace{-4ex}
\scalebox{0.18}[0.18]{\includegraphics[angle=270]{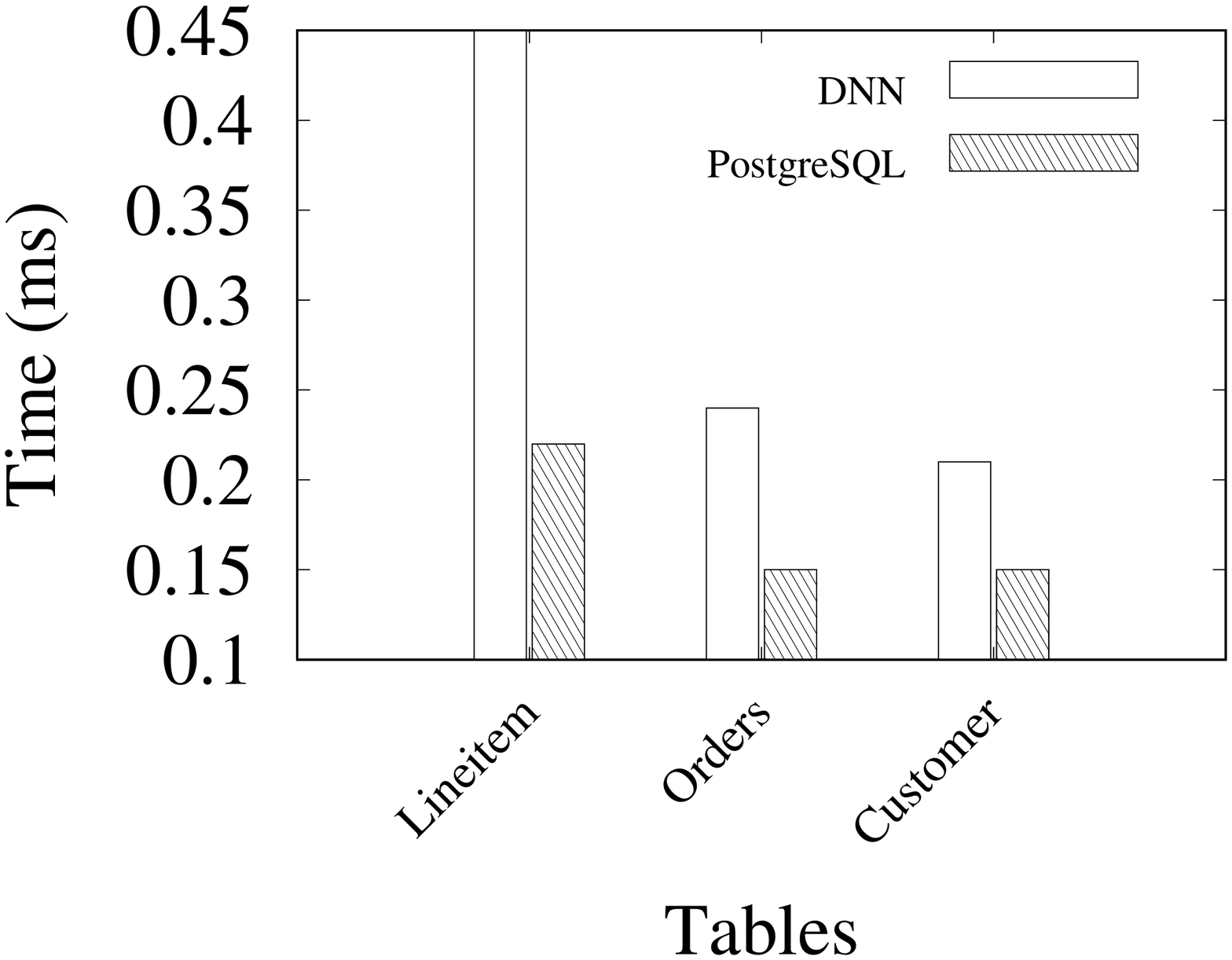}}\label{subfig:select}
}%
\subfigure[][{\small \textsc{Projection} }]{
\hspace{-4ex}
\scalebox{0.18}[0.18]{\includegraphics[angle=270]{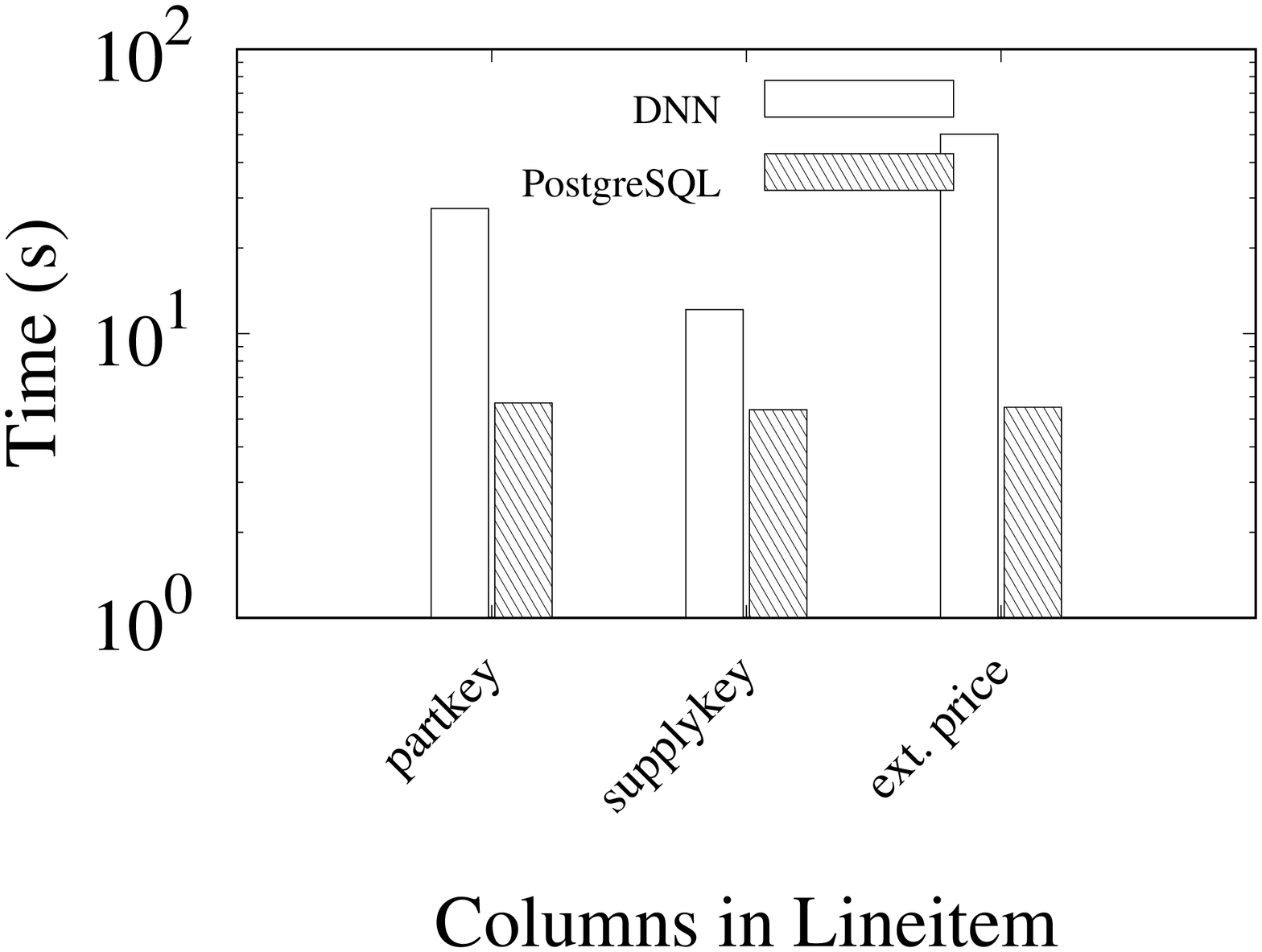}}\label{subfig:project}
}
\subfigure[][{\small \textsc{Range Query}}]{
\hspace{-4ex}
\scalebox{0.18}[0.18]{\includegraphics[angle=270]{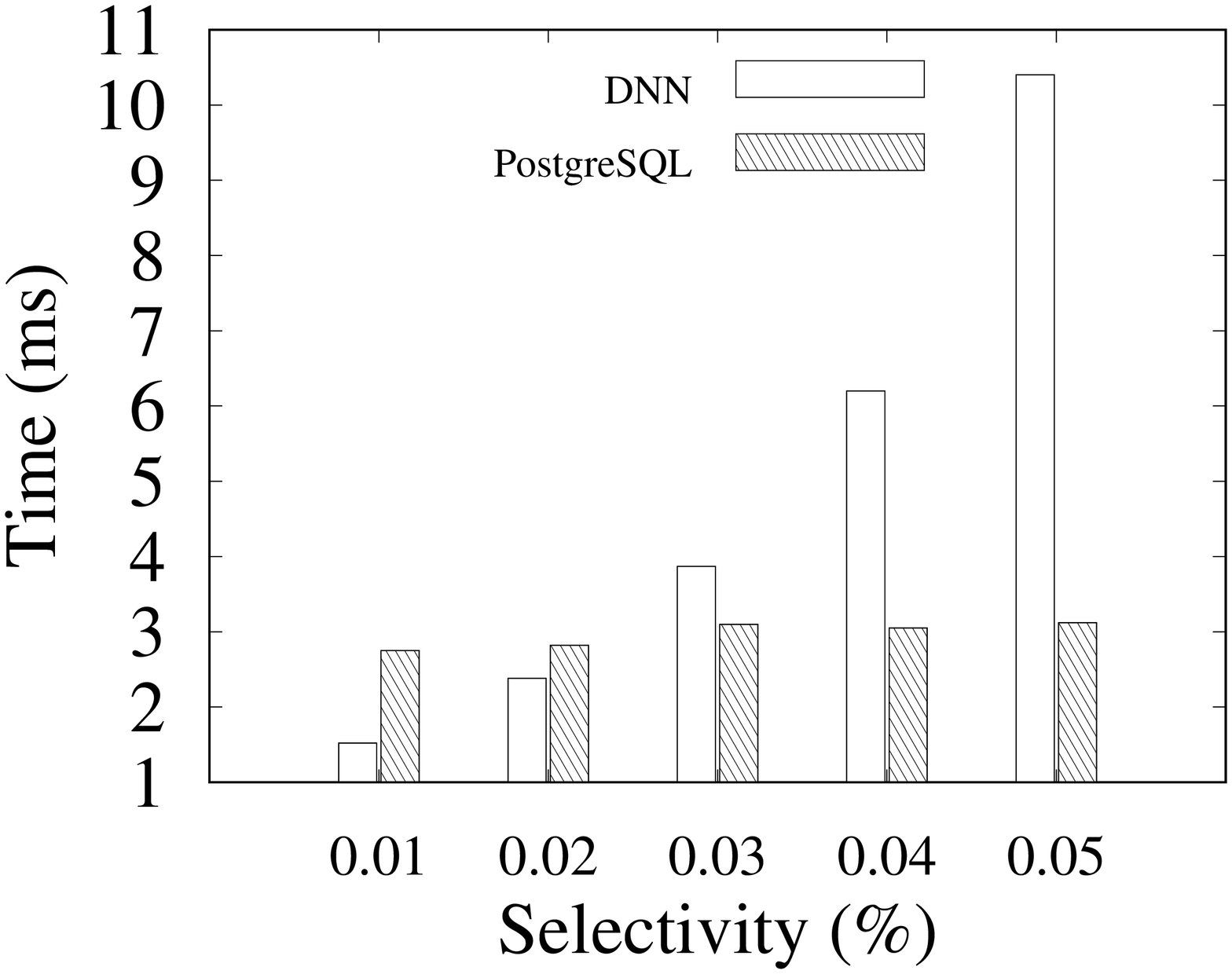}}\label{subfig:range}
}%
\subfigure[][{\small \textsc{Join}\&\textsc{Cartesian Product}}]{
\hspace{-4ex}
\scalebox{0.18}[0.18]{\includegraphics[angle=270]{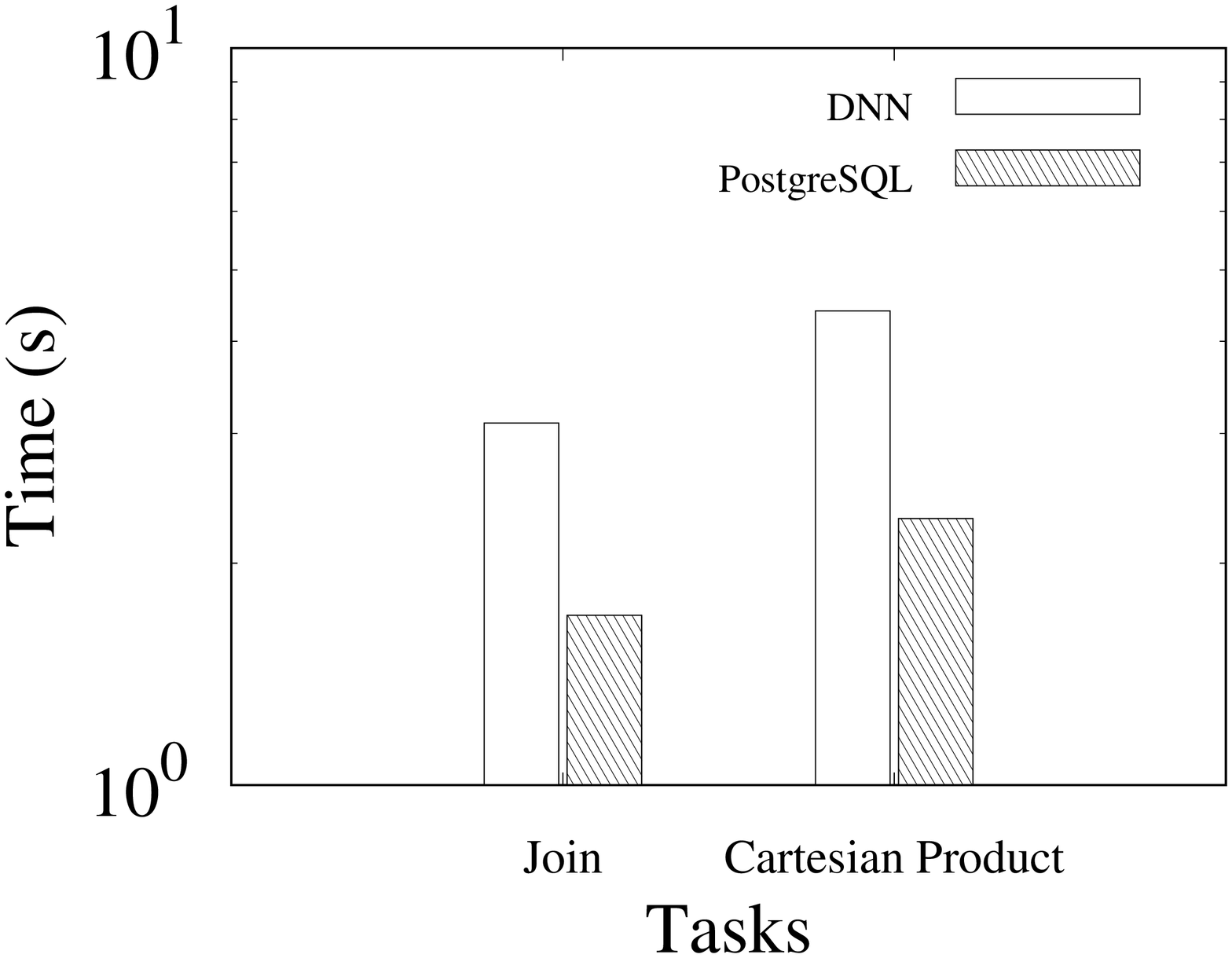}}\label{subfig:join}
}
\caption{\small Evaluation of Primitive SQL Operators over TPC-H-F1.}
\label{fig:capacity}
\end{figure*}

\noindent\textbf{\textsc{Selection}}. We evaluate the time cost to select a tuple with a particular primary key over {\sf TPC-H-F1} data set. As shown in Figure~\ref{subfig:select}, the time cost of retrieving a tuple using PostgreSQL is comparable to that using our DNN-based data storage (i.e., recovering attributes of the tuple from DNNs). From Figure~\ref{subfig:select}, the selection performance varies for different tables. For example, the selection over the Lineitem table takes the longest time. This is because this table's column number is almost twice as that of the other two, and the DNN-based selection needs to access 16 columns, i.e., 16 DNNs, to compute the entire tuple. Nevertheless, the selection operator over our DNN-based data storage incurs low time cost (i.e., less than 0.45 $ms$), and our DNN-based data storage (if adopted over encrypted inputs/outputs) can achieve a higher security level compared with PostgreSQL. 


\noindent\textbf{\textsc{Projection}}. We conduct a \textsc{Projection} operator on attribute ``partkey'', ``supplykey'', and ``extended price'', respectively, over the Lineitem table. As shown in Figure~\ref{subfig:project}, the time costs of the projection on each of these attributes via PostgreSQL stay almost the same, whereas that of our DNN-based data storage exhibits varied performance. For our DNN solution, we return not only the projected attribute values, but also tuple identifiers. Therefore, we use the index in $\mathcal{D}.aux$ to access distinct attribute values (and their tuple counts), and apply DNNs, $DNN_j$ (as discussed in Section \ref{subsec:selection}), to obtain tuple identifiers. In Figure~\ref{subfig:project}, the projection on the ``extended price'' attribute has the highest time cost, since it has the most unique attribute values (i.e., 933,900). From the experimental results, we can see that the number of DNN accesses can dominate the projection efficiency in our DNN-based data storage.





\noindent\textbf{\textsc{Range Query}}. Our DNN-based solution evaluates the \textsc{Range Query}, by first looking up the index of a given attribute in $\mathcal{D}.aux$ to identify the group of tuple identifiers (computed via DNNs). Then, we obtain the entire tuples by feeding tuple identifiers into DNNs for different attributes. Figure~\ref{subfig:range} shows the performance of our design for range queries over the ``total price'' attribute in the Order table, by varying the selectivity from 0.01\% to 0.05\%. When the selectivity value increases, the time costs of both PostgreSQL and DNN approaches increase smoothly. 

Since PostgreSQL built an index over the attribute ``total price'', it can utilize the index to find the first tuple satisfying the range predicate and perform the sequential scan for the remaining range query answers. In contrast, our DNN-based data storage has to compute every tuple satisfying the range predicate through DNNs (i.e., we cannot take advantage of accessing clustered tuples like PostgreSQL). Therefore, our DNN-based data storage requires a higher time cost than PostgreSQL for range queries with high selectivity values. Nonetheless, the time cost of the range query for our DNN-based solution remains low (i.e., less than 10 $ms$). Moreover, it is worth noting that when the selectivity value is small (e.g., 0.01\%), the range query of our DNN-based data storage outperforms that of PostgreSQL. 

We believe that our DNN-based solution (over encrypted inputs/outputs) can provide higher security level, and has the potential to further improve the performance in the multi-core environment that allows computing tuples in parallel.


\noindent\textbf{\textsc{Join \& Cartesian Product}}. Figure~\ref{subfig:join} evaluates the performance of \textsc{Join} and \textsc{Cartesian Product} operators over two tables, Supplier and Nation, of sizes 1.4 $MB$ and 2.2 $KB$, respectively. For the \textsc{Join} operator, we consider the natural join over Supplier and Nation tables, with the join predicates on attribute ``nationkey''. Each supplier tuple is joined with exactly one nation tuple. Our DNN-based data storage takes 3.12 $secs$ to complete the join, whereas PostgreSQL takes 1.7 $secs$. We observe that the major time cost of our DNN solution is in the computation of attribute values for returning join outputs. 

The Cartesian product is expected to be more time-consuming than \textsc{Join}. The extra cost of \textsc{Cartesian Product} mainly comes from returning a 25x larger output compared to \textsc{Join}. The time cost of our DNN-based data storage is 4.4 $secs$ and that of PostgreSQL is 2.7 $secs$. 



\noindent {\bf Summary and Discussions.} Overall, the experiments above show the viability of supporting primitive SQL operators over DNN-based data storage. Although the current implementation is rather naive, as suggested by the performance gap when compared to mature relational database products, our DNN-based data storage in fact trades the computation time for a higher security level (as the majority of the time cost is from computing attribute outputs from DNNs). Also, our DNN-based data storage over encrypted inputs/outputs can enable secure query processing without data decryption (which cannot be achieved by many prior works).

From our experimental results, we also observe many research opportunities that can bring ``DNN-as-a-Database'' to real practice. First, a quantitative study of how different sizes of DNN storage blocks would affect the overall performance would help derive an adaptive solution for encoding data into DNNs to strike a balance among the DNN transformation cost, space overhead, and query performance. Second, although DNN-based query evaluation is CPU intensive, it can significantly reduce the I/O cost. Thus, our design can greatly benefit from a multi-core or CPU/GPU hybrid computing environment. Moreover, instead of simply taking DNNs as storage, we can further develop DNN-based caching, query decomposition/rewrite, and scheduling to improve the overall efficiency of data management via DNNs.

\nop{
data sets: Uniform, Gaussian, Skewed (Zipf), TCP-H?\\

* the number of stored tuples vs. DNN sizes (parameters: dimensionality $d$, DNN parameters such as $L$ and No. of neurons for each hidden layer).\\

* the space cost of the DNN-based data storage vs. the capacity of the DNN-based data storage (plotted chart).\\

* the time cost vs. the data size (sequential scan, tuple existence checking, tuple insertion/deletion/updates, equality/range search, join, union/intersection/set difference).
}

\section{Related Work}
\label{sec:related_work}

\noindent {\bf Neural Networks and Recent Advancements.} Originated from Artificial Neural Networks (ANNs), \textit{deep learning} \cite{Schmidhuber14} is one of the most performant machine learning tools, as examplified by Recurrent Neural Networks (RNN) and Convolutional Neural Networks (CNN)~\cite{DBLP:conf/cvpr/KarpathyTSLSF14,DBLP:conf/nips/KrizhevskySH12}. The recent advancements of Graphics Processing Units (GPUs) and distributed computing techniques have made it feasible to train a model with billions of variables using large-scale training data sets~\cite{DBLP:journals/corr/AzarkhishRLB17,DBLP:journals/corr/McMahanMRA16,DBLP:conf/iri/YanZSC16}. With extensive learning-based solutions developed for solving real-world applications, researchers are now paying more attention to the cross-modality structure~\cite{DBLP:journals/corr/abs-2105-00642}, model explainability~\cite{DBLP:journals/corr/abs-2004-14545}, model robustness~\cite{DBLP:journals/corr/abs-1810-11726}, and security of neural networks~\cite{DBLP:conf/dac/RouhaniRK18}. 


Previous works on neural networks usually incur the prediction/classification inaccuracy, since the trained neural networks are used for predicting/classifying unknown testing data. In our learning-based data storage, however, we treat tabular data as both training and testing data sets. Thus, as long as we can achieve high accuracy of neural networks during the training process and appropriately handle inaccurate cases (e.g., using more neural networks), we can guarantee accurate (e.g., 100\%) data retrieval from neural networks. 

\noindent {\bf Learning-based Data Management.} Recently, there is an emerging trend of applying learning-based methods to traditional database management systems, ranging from architecture design and query optimization to tuning overall performance. Below, we briefly categorize the most related efforts in this direction. 

    \underline{\it Learning-based Database Design:}   Studies~\cite{DBLP:conf/sigmod/DingMYWDLZCGKLK20,DBLP:conf/cidr/IdreosDQAHRLJGL19,DBLP:conf/sigmod/NathanDAK20} show that learning-based indexes can not only reduce the index size but also improve query performance. For example, Kraska et al.~\cite{DBLP:conf/sigmod/KraskaBCDP18} propose that indexes are models, where the B$^+$-tree index can be seen as a model that maps each query key to its page. Techniques like data structure alchemy~\cite{DBLP:conf/cidr/IdreosDQAHRLJGL19} are proposed to automatically recommend and design data structures that best suit the computing workload and hardware setting. The essential idea is to first identify the bottleneck of the total cost and then tweak different knobs in one direction until reaching a pre-defined cost metric or finding the minimal total cost, which is similar to the gradient descent procedure. 
    
    \underline{\it Learning-based Database Configuration:} Machine learning-based approach has been widely applied to a range of database configurations, including knob tuning, index advisor, materialized view advisor, and database partition. For example, researchers utilize learning-based techniques~\cite{DBLP:conf/sigmod/KunjirB20,DBLP:journals/pvldb/LiZLG19,DBLP:conf/sigmod/ZhangLZLXCXWCLR19} to automate knob tuning, which not only achieve higher tuning performance but less tuning time. CDBTune~\cite{DBLP:conf/sigmod/ZhangLZLXCXWCLR19} models database tuning as a sequential decision problem and relies on reinforcement learning to improve tuning performance. QTune~\cite{DBLP:journals/pvldb/LiZLG19} further characterizes query features using deep learning and can achieve finer granularity tuning, e.g., query-level tuning, session-level tuning, and system-level tuning.  Indexes are vital to speed up query execution, and there are some learning-based works that automatically recommend indexes~\cite{DBLP:journals/pvldb/KossmannHJS20,DBLP:conf/sigmod/MaDDS20,DBLP:conf/ideas/SadriGL20}. View materialization usually provides an effective space-for-time trade-off in performance improvement. Yet view selection is an NP-hard problem. Learning based solutions \cite{DBLP:journals/pvldb/KossmannHJS20,DBLP:journals/corr/abs-1903-01363,DBLP:conf/icde/Yuan0FSH20} aim to automatically identify the appropriate views for a given query workload. For example, Han et al. ~\cite{DBLP:conf/icde/Han00S21} propose a deep reinforcement learning method to estimate the benefit of different materialized view candidates and queries for dynamic workloads. Another important aspect of configuration is data partitioning. Instead of heuristically  selecting columns as partition keys (single column mostly), which cannot balance between load balance and access efficiency, recent work by Hilprecht et al.~\cite{DBLP:conf/sigmod/HilprechtBR20} utilizes a reinforcement learning model to explore different partition keys and implements a fully-connected neural network to estimate partition benefits.

    \underline{\it Learning-based Database Optimization:} This line of research aims to utilize machine learning techniques to address the hard problems in database optimization, including cost estimation, join order selection, and end-to-end optimization. Specifically, deep learning based techniques (e.g., CNN~\cite{DBLP:journals/pvldb/DuttWNKNC19}, RNN~\cite{DBLP:journals/pvldb/SunL19}, Mixture Model~\cite{DBLP:conf/sigmod/ParkZM20}) are proposed to estimate the cost and cardinality  to capture the correlations between different columns/tables. For example, an LSTM based work~\cite{DBLP:journals/pvldb/SunL19} learns a representation for each sub-plan with physical operators and predicates and outputs the estimated cardinality and cost simultaneously by using an estimation layer. Given that an SQL query may have an exponential number of possible execution plans, traditional heuristic methods cannot find optimal plans for dozens of tables, and it is costly to use dynamic programming to explore the huge plan space. Thus, prior works propose deep reinforcement learning based methods~\cite{DBLP:conf/sigmod/MarcusP18,DBLP:journals/pvldb/MarcusNMZAKPT19,DBLP:conf/icde/Yu0C020} that automatically select good plans. For example, SkinnerDB~\cite{DBLP:journals/corr/abs-1901-05152} uses a Monte-Carlo tree search based method that can optimize the join order on the fly by trying out different join orders in each time slice.  Learning-based end-to-end optimizers~\cite{DBLP:conf/sigmod/MarcusNMTAK21,DBLP:journals/pvldb/MarcusNMZAKPT19,DBLP:journals/pvldb/WuJAPLQR18} use deep neural networks to optimize SQL queries. For example, Marcus et al.~\cite{DBLP:journals/pvldb/MarcusNMZAKPT19} propose NEO, which uses PostgreSQL’s plan to pre-train the neural network and latency as feedback to train the neural network. In contrast to the rule-based traditional workload prediction methods~\cite{DBLP:conf/sigmod/DasLNK16} in transaction management,  Ma et al.~\cite{DBLP:conf/sigmod/MaAHMPG18} propose a learning-based system that predicts the future trend of different workloads. Besides, recent research~\cite{DBLP:journals/corr/abs-1903-02990} shows that  the learning based transaction scheduling method can balance concurrency and conflict rates using supervised algorithms. 

    \underline{\it Autonomous Database Systems:} There have been research efforts on autonomous database systems from academia~\cite{DBLP:conf/cidr/KraskaABCKLMMN19,DBLP:journals/debu/0001ZL19,DBLP:conf/cidr/PavloAALLMMMPQS17} and industry~\cite{DBLP:journals/pvldb/DasYZVVKGKM15,DBLP:conf/sigmod/DasLNK16,DBLP:journals/pvldb/Li19,DBLP:journals/pvldb/0001ZSYHJLWL21}. Peloton~\cite{DBLP:conf/cidr/PavloAALLMMMPQS17} predicts the workload arrival rate with clustering and RNN and deploys optimization actions like knob tuning and index/materialized view selection. SageDB~\cite{DBLP:journals/pvldb/AhmedBWK20} provides a vision to specialize the database implementation by learning data distributions (CDF models) and designing database components based on the knowledge, e.g., learned index and query scheduling.  Oracle’s autonomous data warehouse~\cite{DBLP:journals/pvldb/DasYZVVKGKM15} optimizes OLAP services with tuning choices like predicate indexing and materialized views. Alibaba’s SDDP~\cite{DBLP:journals/pvldb/Li19} detects database status from historical statistics and provides automatic optimization like hot/cold separation and buffer tuning. 
    OpenGauss~\cite{DBLP:journals/pvldb/0001ZSYHJLWL21} represents the most recent advancements in autonomous database system that integrates a variety of learning-based optimizations to many core components of database systems. 
    

Although extensive research efforts have been devoted to incorporating learning-based methods to various aspects of database systems, most of them are based on a database system that physically stores tables, graphs, or key-value pairs. The essence of our proposed solution is ``DNN-as-a-database'', where all data management are done via the LMU interface, instead of traditional storage media.  




\section{Discussions on DNN-based Graph Storage and Analytics}
\label{sec:graph_discussions}

In this section, we discuss the potential of generalizing our proposed learning-based data storage to other data types such as DNN-based graphs.

\noindent {\bf Graph Data Model.} We first give the definition of the graph data model below.

\begin{definition} \textbf{(Graph, $G$)}
A \textit{graph}, $G$, is represented by a triple $(V(G), E(G), \phi(G))$, where $V(G)$ is a set of vertices $v_i$, each associated with a label $l(v_i)$, $E(G)$ is a set of edges, $e_{ij}$, between vertices $v_i$ and $v_j$, and $\phi(G)$ is a mapping function $V(G)\times V(G) \rightarrow E(G)$.
\label{def:graph}
\end{definition}

The graph model has been widely used in many real applications such as knowledge (RDF) graphs in the Semantic Web, social networks, road networks, and so on.

\noindent{\bf DNN-based Graph Model.} Similar to DNN-based relational tables, for a graph $G$, we will design LMU, $\mathcal{G}$, over $G$, which includes two components, $\mathcal{G}.DNN$ and $\mathcal{G}.aux$. For $\mathcal{G}.DNN$, we consider vertices $v_i \in V(G)$ and their 1-hop neighbors (i.e., vertices adjacent to $v_i$) in a vector $NN(v_i)$ and train DNNs, $DNN_G \sim ((l(v_i)), (v_i, NN(v_i)))$ and $DNN_G^{-1} \sim ((v_i, NN(v_i)), (v_i.id))$. For $\mathcal{G}.aux$, we store auxiliary information such as indexes built over vertex labels $l(v_i)$.

Note that, in some real applications such as the Semantic Web, RDF graphs can be equivalently represented by RDF triples, which can be considered as tuples and alternatively stored via DNN-based relational tables.

\noindent{\bf Queries Over DNN-based Graphs.} We consider two classic graph queries such as graph isomorphism checking and subgraph matching over DNN-based graphs, $DNN_G$, in LMUs. Specifically, we formally define these two DNN-based graph queries as follows.

\begin{definition} \textbf{(Graph Isomorphism Checking Over DNN-Based Graphs, GIC-DNN$_G$)} 
Given $N$ DNN-based graphs $g$ in a graph database and a query graph $q$, \textit{graph isomorphism checking over DNN-based graphs} (GIC-DNN$_G$) retrieves graphs $g$ that are isomorphic to $q$ (denoted as $g \equiv q$).
\label{def:graph_iso}
\end{definition}

\begin{definition} \textbf{(Subgraph Matching Over DNN-Based Graphs, SM-DNN$_G$)} 
Given a DNN-based graph $G$ and a query graph $q$, a \textit{subgraph matching query over DNN-based graphs} (SM-DNN$_G$) retrieves subgraphs $g \subseteq G$ such that $g$ is isomorphic to $q$ (i.e., $g \equiv q$).
\label{def:subgraph_matching}
\end{definition}

 \underline{\it GIC-DNN$_G$:} In order to enable efficient GIC-DNN$_G$ query processing, we use a filter-and-refinement framework that first retrieves graph candidates through a filtering step and then performs the actual isomorphism checking in the refinement step. For the filtering step, we consider labels of each vertex $q_i$ and its 1-hop neighbors $NN(q_i)$ in the query graph $q$ as input and use $DNN_G^{-1} \sim ((q_i, NN(q_i)), (v_i.id))$ in $\mathcal{G}.DNN$ to retrieve candidate vertex IDs $v_i.id$ in graphs $g$. Then, we can combine/join candidate vertices $v_i$ of those vertices $q_i$ in $q$ and obtain candidate subgraphs $g$. Finally, we can refine these candidate subgraphs via actual graph isomorphism checking. 

\underline{\it SM-DNN$_G$:} In the GIC-DNN$_G$ problem, vertices and their 1-hop neighbors in data graph $g$ and query graph $q$ must exactly match with each other. In contrast, for the SM-DNN$_G$ problem, each vertex $q_i$ and its 1-hop neighbors $NN(q_i)$ in query graph $q$ must be a subgraph of a matching vertex $v_i\in V(G)$ and its 1-hop neighbors $NN(v_i)$ in $G$. Thus, in this case, techniques for GIC-DNN$_G$ by using $DNN_G^{-1}$ do not work. In order to resolve this issue, we can design an effective \textit{graph embedding} approach which utilizes DNNs to encode/map each star graph $g$ (containing a vertex $v_i\in V(G)$ and its 1-hop neighbors $NN(v_i)$) in $G$ to a feature vector $F(g)$ (and a tuple identifier $v_i.id$ as well). In other words, we will offline train a variant of DNNs: $DNN_G^{-1} \sim ((v_i, NN(v_i)), (F(g), v_i.id))$. In this DNN variant, we carefully devise an appropriate \textit{loss function} in the DNN such that any subgraph $g'$ of star graphs $g$ satisfies the condition $F(g') \preceq F(g)$, where ``$\preceq$'' is an order predicate between two vectors $F(g')$ and $F(g)$. Next, we will offline train another variant of DNN: $DNN_G \sim ((F(g)), (v_i))$, which can be used for finding matching vertices $v_i$. 

This way, for any query graph $q$, we can obtain a feature vector $F(g(q_i))$ for each vertex $q_i$. By order constraints, we can obtain ranges of features for those vectors $F(g(v_i))$ such that $g(q_i)$ is a subgraph of $g(v_i)$. Then, we can use $DNN_G \sim ((F(g)), (v_i))$ with feature ranges as the input, and retrieve potential vertices (or their IDs). Finally, we can join candidate vertices for each vertex $q_i$ in $q$, and obtain candidate subgraphs $g$ for refinement. 

\section{Discussions on Distributed DNN-based Data Storage and Computing}
\label{sec:distributed_discussions}

In this section, we discuss the potential of generalizing our proposed learning-based data storage to the distributed environment, that is, distributed DNN-based data storage. In contrast to current distributed neural network solutions, which mostly focus on improving parallelism during the training process (e.g., paralleling the stochastic gradient descent for fast convergence), our proposed study does not necessarily involve distributed neural network training. We assume that each DNN model can be trained on one single physical machine. However, the DNN-based LMUs could be too large to be held in one machine and are thus stored distributively over multiple servers.

With the distributed DNN-based data storage, learning-based structures (rather than original data formats) can be safely stored on multiple distributed servers and only be materialized when certain input data or computation workloads are provided. This capability offers data security and privacy preservation, as well as the potential to store arbitrarily large-scale data and perform the corresponding analysis.

To enable efficient DNN-based data analytics in a distributed setting, we can mainly explore two thrusts. First, we will investigate the strategy to dynamically partition the LMU storage across physical machines to best serve the purpose of workload balance and overall query processing throughput. In contrast to traditional cost-model-guided data partitioning, the benefit or overhead of moving DNNs around is not straightforward. We can devise an interpreter that maps the present data access pattern and workload to the reward/loss of a given partitioning decision. We model it as a Partially-Observable Markov Decision Process (POMDP) and solve it with Q-learning~\cite{DBLP:conf/iclr/WangH0DZ21}. This approach would benefit from a continuous learning process, such that it can provide timely re-partitioning guidance when the workload pattern shifts. Our second thrust is to develop time-efficient query evaluation for a given LMU partitioning outcome. The key to the success of this thrust lies in a number of optimization techniques: 1) an update-friendly light-weighted routing scheme (index) to embed the LMU structure, such that given a query, we would be able to identify not only the set of DNNs to access but also where they are located; 2) reinforcement learning empowered algorithms to learn desired execution plans; and 3) highly expressive and efficient runtime function support (e.g., APIs that support user-defined filter/constraints). 

Consider \textsc{Join} processing in a distributed LMU setting as an example. Being a fundamental operator, \textsc{Join} is commonly adopted in analyses over structured relational tables or semi-structured graphs. For example, it can serve as a fundamental building block for graph isomorphism checking or subgraph matching, where the (sub)graph matching query can be expressed as a multi-way join query~\cite{DBLP:journals/corr/abs-1906-11518}. Considering the huge size variance of intermediate results produced from different join orders, we can incorporate the reinforcement learning strategy to learn a proper join sequence that best overlaps the computation and communication. Specifically,  we train the model by exploring a number of execution plans for a set of benchmark queries. By treating each decision of moving data or models as an action, and the efficiency improvement as the reward, our model can learn a proper scheduling strategy. This approach can be generalized to other complex workloads.


\balance

\section{Conclusions}
\label{sec:conclusion}

In this paper, we envision a novel direction of learning-based data storage which utilizes learning-based structures such as DNNs in a so-called learning-based memory unit (LMU) to store, manage, and analyze data. Although extensive research efforts have been devoted to incorporating learning-based methods to various aspects of database systems, most of them are based on a database system that physically stores tables or key-value pairs. The essence of our proposed study is ``DNN-as-a-database'', such that data are implicitly stored in DNNs (in other words, no explicit values of data records could be revealed by a memory/disk scan), which is useful for data security and privacy preserving and has the potential of distributed DNN-based data storage and computing. 

Our proposed learning-based data storage can be applied to other data types such as DNN-based graphs $G$. Similar to DNN-based relational tables, we store vertices $v_i \in V(G)$ and their 1-hop neighbors (i.e., vertices adjacent to $v_i$) in DNNs. We can use such DNN-based graphs to process graph queries like the graph isomorphism checking or subgraph matching, by finding candidate vertices that match with vertices in a given query graph $q$ via DNNs.

The learning-based data storage can be also generalized to other scenarios such as the distributed environment, that is, distributed DNN-based data storage. Given that data are implicitly encoded in the LMU structure, it opens up a series of research opportunities to explore DNN-based locality and workload adaptive DNN model partitioning, as well as fill the gap of interpreting DNN-based data access patterns to the cost estimation of data/model shuffling in a distributed setting. 

In the future, we will explore learning-based data storage via other ML models for heterogeneous data types (e.g., unstructured data, trees, graphs, and images) and scenarios (e.g., IoT devices, distributed environment, etc.) and study its utilities in various real-world applications that demand the data security, privacy preserving, and/or distributed storage/computing in a scalable manner.

\nop{
\section{Future Research Directions}
\label{sec:future_work}

* core techniques of merging two DNNs (without enumerating all data)

* hardware improvement for DNNs

* generalization to other data types such as graphs

* distributed computing (key/value pairs)
}


\bibliographystyle{ACM-Reference-Format}
\bibliography{all}

\end{document}